\documentclass[%
 aip,
 amsmath,amssymb,
 reprint,%
]{revtex4-1}
\usepackage{graphicx}
\usepackage{dcolumn}
\usepackage{bm}

\usepackage[utf8]{inputenc}
\usepackage[T1]{fontenc}
\usepackage{mathptmx}
\usepackage{etoolbox}
\usepackage{physics}
\usepackage[shortlabels]{enumitem}

\usepackage[final]{changes}

\makeatletter
\def\@email#1#2{%
 \endgroup
 \patchcmd{\titleblock@produce}
  {\frontmatter@RRAPformat}
  {\frontmatter@RRAPformat{\produce@RRAP{*#1\href{mailto:#2}{#2}}}\frontmatter@RRAPformat}
  {}{}
}%
\makeatother

\begin{document}
\preprint{AIP/123-QED}
\title{Extracting double-quantum coherence in two-dimensional electronic spectroscopy under pump-probe geometry}
\author{Mao-Rui Cai}
\affiliation{Graduate School of China Academy of Engineering Physics, Beijing 100193,
China}
\author{Xue Zhang}
\affiliation{Graduate School of China Academy of Engineering Physics, Beijing 100193,
China}
\author{Zi-Qian Cheng}
\affiliation{Graduate School of China Academy of Engineering Physics, Beijing 100193,
China}
\author{Teng-Fei Yan}
\affiliation{School of Microelectronics, Shanghai University, Shanghai 200444,
China}
\author{Hui Dong}
\email{hdong@gscaep.ac.cn}

\affiliation{Graduate School of China Academy of Engineering Physics, Beijing 100193,
China}
\begin{abstract}
Two-dimensional electronic spectroscopy (2DES) can be implemented with different geometries, e.g., BOXCARS, collinear and pump-probe geometries. The pump-probe geometry has its advantage of overlapping only two beams and reducing phase cycling steps. However, its applications are typically limited to observing the dynamics with single-quantum coherence and population, leaving the challenge to measure the dynamics of the double-quantum (2Q) coherence, which reflects the many-body interactions. We \replaced{demonstrate }{propose} an experimental technique in 2DES under pump-probe geometry with a designed pulse sequence and the signal processing method to extract 2Q coherence. In the designed pulse sequence with the probe pulse arriving earlier than pump pulses, our measured signal includes the 2Q signal as well as the zero-quantum (0Q) signal. With phase cycling and data processing using causality enforcement, we extract the 2Q signal. The proposal is demonstrated with the rubidium atoms. We observe the collective resonances of two-body dipole-dipole interactions of both $D_{1}$ and $D_{2}$ lines.
\end{abstract}
\maketitle

\section{Introduction}

Two-dimensional (2D) spectroscopy~\citep{mukamel1995book,cho2009book,hamm2011concepts,Jonas2003}, inspired by multi-dimensional nuclear magnetic resonance, stands as a powerful nonlinear spectroscopic approach~\citep{mukamel1995book} for investigating the couplings, structure, and dynamics of systems in both gaseous~\citep{Tian2003,Dai2012,Cundiff2017,Lomsadze2018,Nelson2021} and condensed phases~\citep{Schlau-Cohen2011,Thomas2009}. With advancements in ultrashort pulsed-laser technology, this method has achieved notable success, particularly in the infrared (IR)~\citep{Hochstrasser2000,Tokmakoff2002,Chris2010,Chris2012} and visible regions~\citep{Miller2004,Fleming2005,Fleming2016}, and has been further extended into the ultraviolet (UV)~\citep{Tseng2009,Moran2012}, terahertz (THz)~\citep{Lu2016,Reimann2021}, and hybrid regions~\citep{Thomas2014,Dong2015,Lewis2015,Courtney2015}.

2D spectroscopy is primarily employed for the study of third-order responses, and also has been developed to explore signals of higher orders. The experimental realization of third-order 2D spectroscopy typically involves a sequence of four laser pulses with a well-defined time ordering~\citep{Jonas2003,CundiffMukamel2013}. Among these pulses, three are employed to induce the third-order nonlinear response~\citep{mukamel1995book} of a sample, while the fourth serves as an auxiliary pulse for detection purposes. Specifically, by interacting with the sample, the first pulse generates a coherence upon arrival. Subsequently, the second pulse transforms the coherence into either a double-quantum (2Q)~\citep{Kim2009} or zero-quantum (0Q) coherence, which is then converted into another coherence by the third pulse. In the context of heterodyne detection, the third-order nonlinear signal generated by the final coherence is detected by treating the auxiliary pulse as the local oscillator (LO)~\citep{Tobias2004,Stone2009,Grumstrup2007}. Alternatively, in fluorescence detection, the auxiliary pulse converts the coherence into an excited population, and the fluorescence signal is detected directly in the time domain~\citep{Tian2003}.

Third-order 2D spectroscopy is usually implemented with different geometries, e.g., BOXCARS~\citep{Tobias2004,Stone2009}, \added{collinear~\citep{Tian2003}}, and pump-probe geometries~\citep{Grumstrup2007,Myers2008,Brida2012,Rock2013}. Depending on which intervals between the three excitation pulses are scanned and Fourier transformed, third-order 2D spectroscopy is denoted as one-quantum (1Q) 2D spectroscopy, or 2Q (or 0Q) 2D spectroscopy~\citep{Kim2009,Zhang1999,Fulmer2004,Gao2016,Yu2022}. Third-order 2Q 2D spectroscopy~\citep{Zhang1999,Fulmer2004} characterizes the 2Q coherence between the ground and a doubly excited level, which either naturally exists within an individual or originates from many-body interactions~\citep{Kim2009,Stone2009,Dai2012}. \added{In third-order 2Q 2D spectroscopy experiments, various configurations, including the BOXCARS~\citep{Stone2009, Dai2012, Turner2010} and collinear geometries~\citep{Yu2019}, as well as the innovative approach using frequency combs~\citep{Cundiff2017,Lomsadze2018} have been widely employed and developed.} Whereas, the pump-probe geometry, though alleviates the challenges of implementing 2D spectroscopy, is barely reported in third-order 2Q studies, limited by its phase matching condition. Attempt was tried to measure the fifth-order 2Q coherence using 2D spectroscopy under pump-probe geometry~\citep{Patrick2020}. Yet the associated signals are generally much weaker compared to the third-order ones.

\added{Recently, the measurement of third-order 2Q coherence using 2D IR~\citep{Farrell2022} and 2D electronic~\citep{Armstrong2023} spectroscopies under pump-probe geometry has been realized with a permuted-pump-probe pulse sequence.} In their sequence, the probe pulse arrives at the sample secondly (or firstly)~\citep{Farrell2022,Armstrong2023}, which is different from the traditional pump-probe sequence~\citep{Grumstrup2007,Myers2008,Brida2012,Rock2013} with the probe pulse arriving lastly. with the permuted-pump-probe pulse sequence, 2Q and 0Q signals have the same phase matching direction and thus will be simultaneously detected. By retrieving the imaginary part of the third-order response from the real-valued signal using a time-domain Kramers-Kronig inversion~\citep{Myers2008,hamm2011concepts,Farrell2022}, 2Q and 0Q coherences are separated at opposite quadrants $(+\omega_{1},+\omega_{3})$ and $(-\omega_{1},+\omega_{3})$ on a common spectrum.

In this paper, we \replaced{demonstrate the extraction of 2Q coherence in }{extend this detection technique to} 2Q 2D electronic spectroscopy (2Q 2DES) under pump-probe geometry by detecting the third-order 2Q coherence in rubidium atoms~\citep{Gao2016} with additional data processing method and phase cycling. Our theoretical analysis and experimental methodology rely on utilizing a pulse shaper to transform a single pump pulse into two distinct pump pulses~\citep{Shim2009,Chris2010}. The interval and phases of these two pump pulses are controlled through the acousto-optic modulator (AOM) in the pulse shaper. We find that the effective oscillation frequencies of 2Q and 0Q coherences during the scanned interval $T$ are not always opposite, i.e., the 2Q and 0Q coherences could potentially reside at the same quadrant on the 2D spectrum. Therefore, we employ the phase cycling scheme $(0,0)$ and $(0,\pi/2)$ of the two pump pulses~\citep{Myers2008} to effectively separate the 2Q and 0Q coherences into different spectra, allowing for their independent analysis without mutual interference. In addition, to the best of our knowledge, we have firstly demonstrated the simultaneous observation of 2Q coherences both for an individual rubidium atom~\citep{Gao2016} and for collective resonances arising from dipole-dipole interactions associated with both two $D$ lines~\citep{Yan2022} of rubidium atoms.

\section{Detecting double-quantum coherence}

\subsection{Phase matching condition and pulse sequence}

\begin{figure}[tbph]
\includegraphics{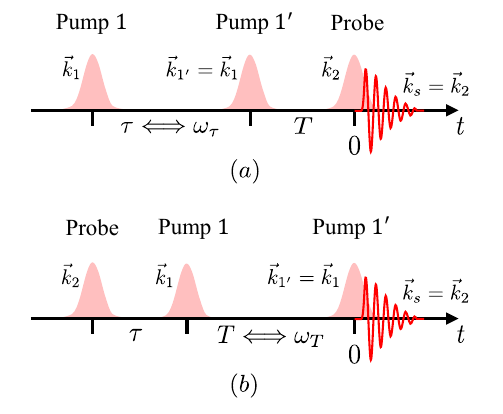}

\caption{Pulse sequences of third-order (a) 1Q 2D spectroscopy and (b) 2Q 2D spectroscopy under pump-probe geometry. The wave vectors of the two pump pulses are $\vec{k}_{1}$ and $\vec{k}_{1'}$ with $\vec{k}_{1'}=\vec{k}_{1}$, and the the wave vector of the probe pulse is $\vec{k}_{2}$. The probe pulse also serves as the LO, and thus third-order signals with wave vector $\vec{k}_{s}=\vec{k}_{2}$ are self-heterodynely detectable. \added{Time zero is set at the moment when the signals are generated.}}
\label{PulseSequence}
\end{figure}

Two-dimensional spectroscopy under pump-probe geometry is a special partial-collinear 2D spectroscopy~\citep{Grumstrup2007,Brida2012}, where two (pump) pulses are collinear and the third (probe) pulse takes a non-collinear direction with respect to these two. Such a 2D spectroscopy has gained exclusive successes in studying the coherent phenomena in several systems in the recent ten years~\citep{Myers2010,Chris2012,Martin2015,Xiang2020,Song2021}. The traditional pulse sequence of the 1Q 2D spectroscopy is shown in Fig.~\ref{PulseSequence}(a). In this sequence, the two pump pulses arrive at the sample before the probe pulse, and the intervals between the three pulses are denoted as $\tau$ and $T$.

\begin{figure}[tbph]
\includegraphics{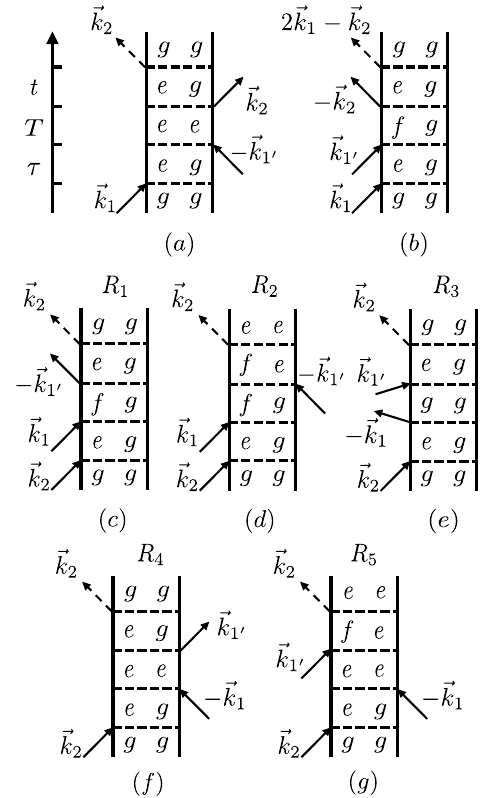}

\caption{Double-sided Feynman diagrams of (a, b) non-rephasing pathways in 1Q 2D spectroscopy, and (c, d) 2Q and (e-g) 0Q pathways in 2Q 2D spectroscopy under pump-probe geometry. In (a) and (b), the probe pulse interacts with the sample lastly, while in (c) to (g) the probe pulse interacts with the sample firstly.}
\label{DoubleSided}
\end{figure}

Under the pump-probe geometry, the third-order signals induced by the three different pulses are typically heterodynely detected in the frequency domain $\omega_{t}$ along phase matching direction $\vec{k}_{s}=\pm\vec{k}_{1}\mp\vec{k}_{1'}+\vec{k_{2}}=\vec{k}_{2}$, where we denote $\vec{k}_{1}$ and $\vec{k}_{1'}$ ($\vec{k}_{1}=\vec{k}_{1'}$) as the wave vectors of the two pump pulses, and $\vec{k_{2}}$ as the wave vector of the probe pulse. Under this geometry, the probe pulse not only interacts with the investigated sample but also serves as the LO~\citep{Shim2009}. In order to fulfill the phase matching condition above, one of the two pump pulses must have its conjugated field interacting with the sample. An example of the detectable pathway for 1Q 2D spectroscopy is shown in Fig.~\ref{DoubleSided}(a), where the first pump pulse induces a coherence $\rho_{eg}$ and the conjugated field of the second pump pulse converts the $\rho_{eg}$ into a population state (0Q coherence) $\rho_{ee}$.

To study the 2Q coherence during $T$, the first two pulses are both used to induce a 2Q coherence (i.e., $\rho_{gg}\rightarrow\rho_{eg}\rightarrow\rho_{fg}$) while the conjugated field of the third pulse converts the 2Q coherence $\rho_{fg}$ into a coherence $\rho_{eg}$, as shown by the pathway Fig.~\ref{DoubleSided}(b). Here, we consider a general three-level system ($\ket{g}\leftrightarrow\ket{e}\leftrightarrow\ket{f}$). The pathway in Fig.~\ref{DoubleSided}(b) corresponds to the traditional pulse sequence of 1Q 2D spectroscopy where the probe pulse arrives lastly, and it suggests that the 2Q signal will emit along $2\vec{k}_{1}-\vec{k}_{2}$. Such a direction is non-collinear with the probe pulse that the 2Q signal cannot be self-heterodynely measured. In order to also detect the 2Q signal, the probe pulse is designed to arrive at the sample firstly or secondly~\citep{Farrell2022}.

We focus on the pulse sequence with the probe pulse arriving firstly in our current study, and we still denote the intervals between the three pulses as $\tau$ and $T$, as shown in Fig.~\ref{PulseSequence}(b). With the probe pulse arriving firstly, the probe pulse and the first pump pulse are used to induce a 2Q coherence $\rho_{fg}$ during $T$, and the conjugated field of the second pump pulse converts the 2Q coherence $\rho_{fg}$ into a coherence $\rho_{eg}$ or $\rho_{fe}$, as presented by the 2Q pathways ($R_{1}$ and $R_{2}$) in Fig.~\ref{DoubleSided}(c) and (d). In these two cases, the 2Q signals have phase matching direction $\vec{k}_{s}=\vec{k_{2}}+\vec{k}_{1}-\vec{k}_{1'}=\vec{k}_{2}$, and thus they are self-heterodynely detectable. However, despite the 2Q signals, 0Q signals with the same phase matching direction will also be measured. We list all detectable 0Q pathways ($R_{3}$, $R_{4}$, and $R_{5}$) in Fig.~\ref{DoubleSided}(e-g). Unlike in $R_{1}$ and $R_{2}$ where the conjugated field of the second pump pulse interacts with the sample, in the 0Q pathways, the conjugated field of the first pump pulse contributes. In specific, after the probe pulse inducing a coherence $\rho_{eg}$, the conjugated field of the first pump converts $\rho_{eg}$ into a 0Q coherence $\rho_{gg}$ or $\rho_{ee}$ during $T$. Then, the second pump pulse induces another coherence $\rho_{eg}$ or $\rho_{fe}$.

Additionally, in the case of the probe pulse arriving secondly, there are also two 2Q and three 0Q pathways self-heterodynely detectable, which are given in the supplementary material. Note that in 2D spectroscopy experiments under pump-probe geometry, the interval between the pump pulses~\citep{Shim2009,Chris2010,Brida2012} and the time ordering of the probe pulse are typically controlled independently. Letting the probe pulse arrive secondly (i.e., inserting the probe pulse between the two pump pulses) will mix up the uncertainties of time control methods.

\subsection{Phase accumulation in heterodyne detection\label{subsec:Phase-accumulation}}

In the conventional 1Q 2D spectroscopy, the signal field $E_{s}(\tau,T,\omega_{t})$ induced by the three different pulses is emitted promptly after the probe pulse, as illustrated in Fig.~\ref{PulseSequence}(a). This signal field is contemporaneously detected by the spectrometer alongside the probe pulse (i.e., the LO), and the heterodynely measured signal is expressed as $S(\tau,T,\omega_{t})=2\mathrm{Re}\left[E_{s}(\tau,T,\omega_{t})E_{\mathrm{LO}}^{*}(\omega_{t})\right]$, where $E_{\mathrm{LO}}(\omega_{t})=E_{2}(\omega_{t})$ is the field of the probe pulse.

However, in the 2Q 2D spectroscopy depicted in Fig.~\ref{PulseSequence}(b), the third-order signal is exclusively emitted upon the arrival of the second pump pulse. In such a scenario, the signal field lags behind the probe pulse by an interval $\tau+T$, causing the probe pulse to accumulate an additional phase $\exp[-i\omega_{t}(\tau+T)]$ through propagation before acting as the LO~\citep{Tobias2004,Farrell2022}. Taking into account this phase accumulation, the measured signal is now expressed as 
\begin{equation}
S(\tau,T,\omega_{t})=2\mathrm{Re}\left[E_{s}(\tau,T,\omega_{t})E_{2}^{*}(\omega_{t})e^{i\omega_{t}(\tau+T)}\right],\label{eq:hetero-sig}
\end{equation}
noticing that $E_{\mathrm{LO}}(\omega_{t})=E_{2}(\omega_{t})e^{-i\omega_{t}(\tau+T)}$ now.

\subsection{Rotating frame\label{subsec:Rotating-frame}}

2D spectrum $S(\tau,\omega_{T},\omega_{t})$ is obtained by sampling and Fourier transforming $S(\tau,T,\omega_{t})$ with respect to $T$. To reduce the sampling rate required by the Nyquist--Shannon sampling theorem, a rotating frame~\citep{Shim2009} is applied to the first pump pulse to reduce the effective oscillation frequency of the 2Q signal during interval $T$. This is experimentally achieved by setting an additional phase factor $\exp[i\omega_{\mathrm{rf}}T]$ to the first pump pulse using the AOM in the pulse shaper~\citep{Shim2009,Chris2010}. Here, $\omega_{\mathrm{rf}}$ is the rotating frame frequency.

For 2Q pathways $R_{1}$ and $R_{2}$, the 2Q coherence $\rho_{fg}$ oscillates with frequency $\omega_{fg}$ (i.e., the eigenfrequency of $\ket{f}$) during interval $T$, leading to an oscillation of the third-order response as
\begin{align}
R_{1}(\tau,T,\omega_{eg}) & \propto e^{-i\omega_{fg}T},\\
R_{2}(\tau,T,\omega_{fe}) & \propto e^{-i\omega_{fg}T}.
\end{align}
Here, we have set $\omega_{t}=\omega_{eg}$ (i.e., the eigenfrequency of $\ket{e}$) for pathway $R_{1}$, and $\omega_{t}=\omega_{fe}$ for pathway $R_{2}$. Applying the rotating frame phase factor $\exp[i\omega_{\mathrm{rf}}T]$ to the field of the first pump pulse $E_{1}$, the oscillations of 2Q signal fields are modified as
\begin{align}
E_{R_{1}}(\tau,T,\omega_{eg}) & \propto R_{1}E_{1}E_{1'}^{*}E_{2}\propto e^{-i(\omega_{fg}-\omega_{\mathrm{rf}})T},\\
E_{R_{2}}(\tau,T,\omega_{fe}) & \propto R_{2}E_{1}E_{1'}^{*}E_{2}\propto e^{-i(\omega_{fg}-\omega_{\mathrm{rf}})T}.
\end{align}
Furthermore, taking into account the accumulated phase $\exp[i\omega_{t}(\tau+T)]$ discussed in Sec.~\ref{subsec:Phase-accumulation}, the products of 2Q signal fields and the LO field have oscillations
\begin{align}
E_{R_{1}}(\tau,T,\omega_{eg}) & E_{2}^{*}(\omega_{eg})e^{i\omega_{eg}(\tau+T)}\propto e^{-i(\omega_{fe}-\omega_{\mathrm{rf}})T},\\
E_{R_{2}}(\tau,T,\omega_{fe}) & E_{2}^{*}(\omega_{fe})e^{i\omega_{fe}(\tau+T)}\propto e^{-i(\omega_{eg}-\omega_{\mathrm{rf}})T}.
\end{align}
As a result, 2Q pathways $R_{1}$ and $R_{2}$ manifest themselves at $(\omega_{eg},\omega_{T}^{\mathrm{eff}}=\omega_{fe}-\omega_{\mathrm{rf}})$ and $(\omega_{fe},\omega_{T}^{\mathrm{eff}}=\omega_{eg}-\omega_{\mathrm{rf}})$ respectively on 2D spectrum.

The same operation should also be applied to 0Q pathways $R_{3}$, $R_{4}$, and $R_{5}$. We directly give the results as
\begin{align}
E_{R_{3}}(\tau,T,\omega_{eg}) & E_{2}^{*}(\omega_{eg})e^{i\omega_{eg}(\tau+T)}\propto e^{-i(\omega_{\mathrm{rf}}-\omega_{eg})T},\\
E_{R_{4}}(\tau,T,\omega_{eg}) & E_{2}^{*}(\omega_{eg})e^{i\omega_{eg}(\tau+T)}\propto e^{-i(\omega_{\mathrm{rf}}-\omega_{eg})T},\\
E_{R_{5}}(\tau,T,\omega_{fe}) & E_{2}^{*}(\omega_{fe})e^{i\omega_{fe}(\tau+T)}\propto e^{-i(\omega_{\mathrm{rf}}-\omega_{fe})T}.
\end{align}
Because the phase of $E_{1}^{*}$ is inherited by 0Q signal fields, the rotating frame phase factor is $\exp[-i\omega_{\mathrm{rf}}T]$ for 0Q pathways. 0Q pathways $R_{3}$ and $R_{4}$ both generate peaks at $(\omega_{eg},\omega_{T}^{\mathrm{eff}}=\omega_{\mathrm{rf}}-\omega_{eg})$, and the pathway $R_{5}$ generates a peak at $(\omega_{fe},\omega_{T}^{\mathrm{eff}}=\omega_{\mathrm{rf}}-\omega_{fe})$. Similar analysis can be performed for all the pathways presented in the supplementary materials.

\begin{figure}
\includegraphics{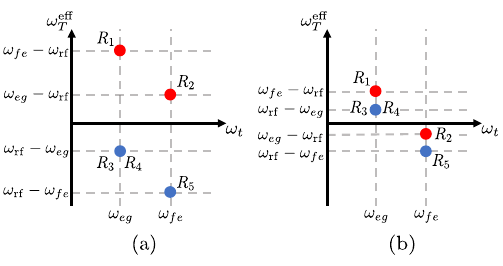}

\caption{Schematic representation of peak positions for 2Q and 0Q signals. The red dots denote 2Q pathways, and the blue dots denote 0Q pathways. The y axis denotes the effective oscillation frequency $\omega_{T}^{\mathrm{eff}}$ during $T$. In (a), the rotating frame $\omega_{\mathrm{rf}}$ is set smaller than both $\omega_{eg}$ and $\omega_{fe}$, and 2Q and 0Q pathways occupy the opposite quadrant along the axis of $\omega_{T}^{\mathrm{eff}}$. In (b), $\omega_{\mathrm{rf}}$ is set between $\omega_{eg}$ and $\omega_{fe}$, and 2Q and 0Q pathways could potentially reside at both quadrants.\label{fig:eff2D}}
\end{figure}

The setting of the rotating frame determines the signs of effective oscillation frequencies of 2Q and 0Q signals during $T$. For instance, if $\omega_{\mathrm{rf}}<\{\omega_{eg},\omega_{fe}\}$, 2Q pathways $R_{1}$ and $R_{2}$ exhibit positive effective oscillation frequencies, whereas 0Q pathways $R_{3}$, $R_{4}$, and $R_{5}$ display negative ones. This situation is illustrated in Fig.~\ref{fig:eff2D}(a). Conversely, if $\omega_{\mathrm{rf}}>\{\omega_{eg},\omega_{fe}\}$, the effective oscillation frequencies of 2Q pathways $R_{1}$ and $R_{2}$ become negative, while those of 0Q pathways $R_{3}$, $R_{4}$, and $R_{5}$ turn positive. In these two cases, 2Q and 0Q signals can be independently analyzed in opposite quadrants along the axis of $\omega_{T}^{\mathrm{eff}}$. However, when $\omega_{\mathrm{rf}}$ is set between $\omega_{eg}$ and $\omega_{fe}$, both 2Q and 0Q pathways could potentially reside at either quadrants. A depiction of such a situation is provided in Fig.~\ref{fig:eff2D}(b) where we have assumed that $\omega_{fe}>\omega_{eg}$. The analogous conclusion holds the same if $\omega_{fe}<\omega_{eg}$.

Moreover, as presented in Eq.~(\ref{eq:hetero-sig}), the heterodyne measurement is limited to capturing only the real part of the production of the signal field and the LO field. Following Fourier transformation with respect to time $T$, these real-valued 2Q and 0Q signals exhibit both positive and negative frequency components on the axis of $\omega_{T}^{\mathrm{eff}}$, which further mix up 2Q and 0Q coherences on the 2D spectrum. To eliminate the influence of 0Q coherences, we apply techniques of phase cycling and causality enforcement in the time $t$ domain. Detailed explanations of these methods are presented in Sec.~\ref{subsec:Enforcing-causality}.

\section{Extracting double-quantum coherence}

\subsection{Phase cycling to remove other contributions\label{subsec:Phase-cycling}}

In addition to the third-order 2Q and 0Q signals induced by the three different pulses, the perturbed free induction decay (PFID)~\citep{Hamm1995,Patrick2023} $S_{\mathrm{PFID}}(\tau,T,\omega_{t})$ may arise when the sample interacts with the probe pulse once and one of the two pump pulses twice~\citep{Shim2009}. This undesired signal is also self-heterodynely detected, as it meets the phase matching condition $\vec{k}_{s}=\vec{k_{2}}\pm\vec{k}_{a}\mp\vec{k}_{a}=\vec{k}_{2}$ (where $a=1,1'$).

By applying the technique of phase cycling~\citep{Tian2003,Shim2009}, we can eliminate $S_{\mathrm{PFID}}(\tau,T,\omega_{t})$ from the total signal. Specifically, two experiments with phase arrangements $(\phi_{1}=0,\phi_{1'}=0)$ and $(\phi_{1}=0,\phi_{1'}=\pi)$ are conducted shot-to-shot, where $\phi_{1}$ and $\phi_{1'}$ are correspondingly the relative phases (with respect to the original pump pulse before the pulse shaper) of the first and second pump pulses. The measured signals are then linearly combined through subtraction $S(\tau,T,\omega_{t};\phi_{1}=0,\phi_{1'}=0)-S(\tau,T,\omega_{t};\phi_{1}=0,\phi_{1'}=\pi)$, in which $S_{\mathrm{PFID}}(\tau,T,\omega_{t})$ is eliminated since $E_{\mathrm{\mathrm{PFID}}}(\tau,T,\omega_{t})\propto E_{2}\left|E_{a}\right|^{2}$ is independent of the phases of the two pump pulses. In this subtraction, 2Q and 0Q signals $S_{R_{j}}(\tau,T,\omega_{t})$ are enhanced because $E_{R_{l}}(\tau,T,\omega_{t};\phi_{1},\phi_{1'})\propto e^{i(\phi_{1}-\phi_{1'})}$ ($l=1,2$) for 2Q pathways, and $E_{R_{k}}(\tau,T,\omega_{t};\phi_{1},\phi_{1'})\propto e^{-i(\phi_{1}-\phi_{1'})}$ ($k=3,4,5$) for 0Q pathways.

Moreover, the four-frame cycling scheme $(\phi_{1}=0,\phi_{1'}=0)$, $(\phi_{1}=0,\phi_{1'}=\pi)$, $(\phi_{1}=\pi,\phi_{1'}=\pi)$, and $(\phi_{1}=\pi,\phi_{1'}=0)$ could be applied to further eliminate the influence from the scatter of the pump pulses~\citep{Shim2009,hamm2011concepts}.

\subsection{Enforcing causality to extract 2Q signal \label{subsec:Enforcing-causality}}

\added{The extraction of 2Q signal fields is conducted using phase cycling scheme $(\phi_{1}=0,\phi_{1'}=0)$ and $(\phi_{1}=0,\phi_{1'}=\pi/2)$ through subtraction $E_s(\tau,T,\omega_{t};0,0) + i E_s(\tau,T,\omega_{t};0,\pi/2)$. However, the phase information of the complex-valued field $E_s(\tau,T,\omega_{t})$ is hidden in the real-valued signal $S(\tau,T,\omega_{t})$ as presented in Eq.~(\ref{eq:hetero-sig}). The complete field $E_s$ is retrieved using the technique of causality enforcement in the time $t$ domain~\citep{Myers2008,hamm2011concepts,Farrell2022}. The detailed procedure is as follows.}

\added{First, third-order signal $S(\tau,T,\omega_{t})$ is detected by a spectrometer in the frequency $\omega_t$ domain. Then, $S(\tau,T,\omega_{t})$ is inversely Fourier transformed with respect to $\omega_t$ to obtain the time domain signal $S_\mathrm{iFFT}(\tau,T,t') = \mathcal{F}_{\omega_t\rightarrow t'}^{-1} \left[S(\tau,T,\omega_{t})\right] = A(\tau, T, t') + A^*(\tau, T, -t')$ with
\begin{equation}
    A(\tau, T, t') = \mathcal{F}_{\omega_t\rightarrow t'}^{-1} \left[E_{s}(\tau,T,\omega_{t})E_{2}^{*}(\omega_{t})e^{i\omega_{t}(\tau+T)}\right],
\label{iFFT-rawSig}
\end{equation}
where $\mathcal{F}_{\omega_t\rightarrow t'}^{-1} \left[g(\omega_t)\right] = \int_{-\infty}^{\infty} g(\omega_t) e^{-i \omega_t t'} d\omega_t / 2\pi$ is the inverse Fourier transformation of an arbitrary function $g(\omega_t)$ in the frequency $\omega_t$ domain. In this context, we denote the time axis after the inverse Fourier transformation as $t'$. Due to the phase factor $\exp{i \omega_t (\tau + T)}$, the time axis $t'$ is shifted in relation to the original time axis $t$ by $t' = t + \tau + T$ in Eq.~(\ref{iFFT-rawSig}).}

\added{Generally, the bandwidth of the probe pulse spectrum $E_2(\omega_t)$ is significantly broader than that of the emission signal spectrum $E_{s}(\tau,T,\omega_{t})$. Thus, the assumption is made that $E_2(\omega_t) = E_2$ is uniformly distributed within the bandwidth of $E_{s}(\tau,T,\omega_{t})$. Consequently, Eq.~(\ref{iFFT-rawSig}) is modified as
\begin{equation}
\begin{aligned}
    A(\tau, T, t') &= E_{2}^{*}\mathcal{F}_{\omega_t\rightarrow t'}^{-1} \left[E_{s}(\tau,T,\omega_{t})e^{i\omega_{t}(\tau+T)}\right] \\
    &= E_2^* E_{s}(\tau,T,t'-\tau-T) \\
    &= E_2^* E_{s}(\tau,T,t).
\end{aligned}
\end{equation}
Due to causality, the signal field $E_s(\tau,T,t)$ is zero for $t < 0$, implying that $A(\tau, T, t')$ is zero for $t' < \tau + T$. However, the time domain signal $S_\mathrm{iFFT}(\tau,T,t')$ does not fulfill the causality requirement since $A^*(\tau, T, -t')$ is non-zero for $t'<-\tau-T$. Thus, we enforce the causality of $S_\mathrm{iFFT}(\tau,T,t')$ by assigning all the values within $t'<\tau+T$ to zero, i.e., multiplying $S_\mathrm{iFFT}(\tau,T,t')$ by the Heaviside step function $\theta(t'-\tau-T)$, $S'(\tau,T,t')=\theta(t'-\tau-T)S_\mathrm{iFFT}(\tau,T,t')$. The causality-enforced signal in the time domain is proportional to the signal field 
\begin{equation}
    S'(\tau,T,t') = A(\tau, T, t') = E_2^* E_{s}(\tau,T,t'-\tau-T),
\end{equation}
and its Fourier transformation with respect to $t'$ is correspondingly given as
\begin{equation}
    S'(\tau, T, \omega_t) = \mathcal{F}_{t'\rightarrow \omega_t}\left[S'(\tau,T,t')\right]=E_2^* E_{s}(\tau,T,\omega_t) e^{i\omega_t(\tau + T)},
\end{equation}
distinct from the measured signal in Eq.~(\ref{eq:hetero-sig}).
}

After the causality enforcement, the resulting signal $S'(\tau,T,\added{\omega_t})$ becomes complex, with 2Q signals located at $(\omega_{eg},\omega_{T}^{\mathrm{eff}}=\omega_{fe}-\omega_{\mathrm{rf}})$ and $(\omega_{fe},\omega_{T}^{\mathrm{eff}}=\omega_{eg}-\omega_{\mathrm{rf}})$, and 0Q signals located at $(\omega_{eg},\omega_{T}^{\mathrm{eff}}=\omega_{\mathrm{rf}}-\omega_{eg})$ and $(\omega_{fe},\omega_{T}^{\mathrm{eff}}=\omega_{\mathrm{rf}}-\omega_{fe})$, as shown in Fig.~\ref{fig:eff2D}.

To complete the extraction, we apply the phase cycling scheme $(\phi_{1}=0,\phi_{1'}=0)$ and $(\phi_{1}=0,\phi_{1'}=\pi/2)$ to separate 2Q and 0Q signals with the relations
\begin{subequations}
\begin{align}
S_{2Q}^{\mathrm{sep}}(\tau,T,\added{\omega_t})&=S'(\tau,T,\added{\omega_t};0,0)\added{+}iS'(\tau,T,\added{\omega_t};0,\pi/2),\\
S_{0Q}^{\mathrm{sep}}(\tau,T,\added{\omega_t})&=S'(\tau,T,\added{\omega_t};0,0)\added{-}iS'(\tau,T,\added{\omega_t};0,\pi/2).
\end{align}
\label{sep-sig}%
\end{subequations} 
This separation is conducted to avoid the possibility that 2Q and 0Q signals reside at both quadrants along the axis of $\omega_{T}^{\mathrm{eff}}$ on the 2D spectrum, which is presented in Fig.~\ref{fig:eff2D}(b).

\subsection{Frequency shift retrieval}

With the signals in Eq.~(\ref{sep-sig}), 2Q and 0Q 2D spectra are obtained by Fourier transforming $S_{2Q}^{\mathrm{sep}}(\tau,T,\added{\omega_t})$ and $S_{0Q}^{\mathrm{sep}}(\tau,T,\added{\omega_t})$ with respect to $T$. However, due to the phase accumulation represented by $\exp[i\omega_{t}(\tau+T)]$ and the rotating frame $\exp[\pm i\omega_{\mathrm{rf}}T]$, 2Q and 0Q signals in Eq.~(\ref{sep-sig}) are frequency modulated, which has been described in detail in Sec.~\ref{subsec:Rotating-frame}. Authentic spectra are obtained once the frequency shifts are retrieved.

To compensate for the shifts, we firstly Fourier transform the 2Q and 0Q signals in Eq.~(\ref{sep-sig}) with respect to both $T$ and $t$ to derive $S_{\mathrm{2Q}}^{\mathrm{sep}}(\tau,\omega_{T}^{\mathrm{eff}},\omega_{t})$ and $S_{\mathrm{0Q}}^{\mathrm{sep}}(\tau,\omega_{T}^{\mathrm{eff}},\omega_{t})$ respectively. Then the frequency values on the axis of $\omega_{T}^{\mathrm{eff}}$ for $S_{\mathrm{2Q}}^{\mathrm{sep}}(\tau,\omega_{T}^{\mathrm{eff}},\omega_{t})$ are all added by $\omega_{t}+\omega_{\mathrm{rf}}$ to obtain the frequency shift retrieved 2Q 2D spectra $S_{\mathrm{2Q}}^{\mathrm{rtrv}}(\tau,\omega_{T},\omega_{t})$. This operation shifts the effective oscillation frequencies during $T$ back to the authentic values. Similarly, the frequency values on the axis of $\omega_{T}^{\mathrm{eff}}$ for $S_{\mathrm{0Q}}^{\mathrm{sep}}(\tau,\omega_{T}^{\mathrm{eff}},\omega_{t})$ are all added by $\omega_{t}-\omega_{\mathrm{rf}}$ to obtain the frequency shift retrieved 0Q 2D spectra $S_{\mathrm{0Q}}^{\mathrm{rtrv}}(\tau,\omega_{T},\omega_{t})$. Finally, these two frequency-shift-retrieved spectra should be multiplied by an additional phase factor $\exp[i\omega_{t}\tau]$ to compensate for the phase accumulation of the probe pulse during $\tau$.

\section{Experiment demonstration}

In this section, we demonstrate the detection and extraction scheme of the 2Q electronic coherence in experiment with rubidium atoms ($^{87}\mathrm{Rb}$) using the 2Q 2DES. 
\begin{figure}[tb]
\includegraphics{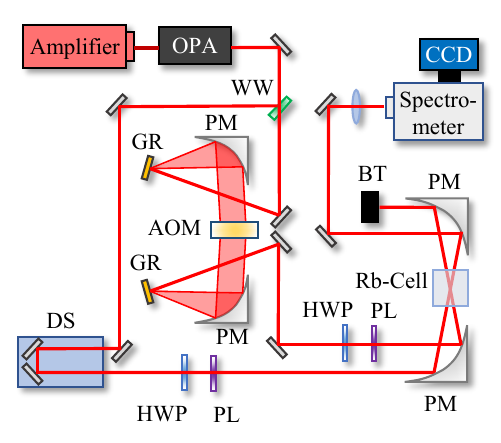}

\caption{The experimental setup of the 2Q 2DES under pump-probe geometry. This setup is identical to the typical 1Q 2DES, whereas the difference is the pulse sequence, as presented in Fig.~\ref{PulseSequence}. Amplifier: Ti:Sapphire amplifier laser system; OPA: optical parametric amplifier; WW: wedged window; PM: parabolic mirror; GR: optical grating; AOM: acousto-optic modulator; DS: delay stage; HWP: half-wave plate; PL: polarizer; BT: beam trap; CCD: charge coupled device camera. The sample is a rubidium ($^{87}\mathrm{Rb}$) vapor cell containing a 450 torr Nitrogen buffer gas. The cell is heated at 120 $^{\circ}\mathrm{C}$ during measurement.}
\label{Setup}
\end{figure}

\subsection{Experimental setup}

The experiment setup is presented in Fig.~\ref{Setup}. A Ti:Sapphire amplifier laser system generates laser pulses (800 nm, 30.8 fs) at a 1 kHz repetition rate. The central wavelength of the laser pulses is shifted using an optical parametric amplifier (OPA). The laser beam after the OPA is divided by a wedged window into the pump and the probe beams. A pulse shaper consisting of an AOM, two parabolic mirrors, and two gratings with 4-f geometry converts one pump pulse into two pump pulses with interval $T$. The interval $\tau$ between the probe and the first pump pulse is controlled by a delay stage on the probe arm. The pump and the probe beams are focused on the sample (a $^{87}\mathrm{Rb}$ vapor cell) using a parabolic mirror. After the sample, the spectrum of the probe pulse is detected by a charge coupled device (CCD) camera connected to a spectrometer. Note that the polarizations of the pump and probe pulses are set horizontal with half-wave plates and polarizers before hitting the sample. In addition, $\tau$ is kept zero in our experiment.

The vapor cell is a cube with an edge length of 15\:mm. During the measurement, the cell is heated at 120 $^{\circ}\mathrm{C}$ with a corresponding atomic density of around $1.65\times10^{13}\:\mathrm{cm^{-3}}$ of $^{87}\mathrm{Rb}$ atoms~\citep{Steck}. In addition, the cell contains a 450 torr Nitrogen buffer gas which broadens the resonance linewidths of $^{87}\mathrm{Rb}$ through collision.

The wavelength of the output pulses from the OPA centers at 783 nm which adjoins the $D_{2}$ transition ($5 ^{2}S_{1/2} \rightarrow 5 ^{2}P_{3/2}$) frequency of $2\pi\times384.2\:\mathrm{THz}$ (780 nm) of $^{87}\mathrm{Rb}$ atoms~\citep{Steck}. The pulse spectrum has a bandwidth (FWHM) of around 24 nm, meaning that not only the $D_{2}$ transition, but also the $D_{1}$ transition ($5 ^{2}S_{1/2} \rightarrow 5 ^{2}P_{1/2}$) at a frequency of $2\pi\times377.1\:\mathrm{THz}$ (795 nm) and the excited states transition ($5 ^{2}P_{3/2} \rightarrow 5 ^{2}D$) at a frequency of $2\pi\times386.3\:\mathrm{THz}$ (776 nm) could be simultaneously induced.

\subsection{Experimental results}

To demonstrate the extraction of 2Q coherence, we utilize a phase cycling scheme $(\phi_{1}=0,\phi_{1'}=0)$, $(\phi_{1}=0,\phi_{1'}=\pi/2)$, $(\phi_{1}=0,\phi_{1'}=\pi)$, and $(\phi_{1}=0,\phi_{1'}=3\pi/2)$, and the corresponding spectra (denoted as $S_{1}$, $S_{2}$, $S_{3}$, and $S_{4}$) are detected shot-to-shot. These spectra are combined by subtractions to generate two spectra $S(0,T,\omega_{t};0,0)=S_{1}(0,T,\omega_{t})-S_{3}(0,T,\omega_{t})$ and $S(0,T,\omega_{t};0,\pi/2)=S_{2}(0,T,\omega_{t})-S_{4}(0,T,\omega_{t})$ which correspond to spectra in cycling scheme $(\phi_{1}=0,\phi_{1'}=0)$ and $(\phi_{1}=0,\phi_{1'}=\pi/2)$ with the PFID signal eliminated. The rotating frame is set at 787 nm, corresponding to a frequency of $\omega_{\mathrm{rf}}/2\pi=380.9\:\mathrm{THz}$, and $T$ is scanned from 0 to 14.4 ps with a step size of 80 fs.
\begin{figure}[tbph]
\includegraphics{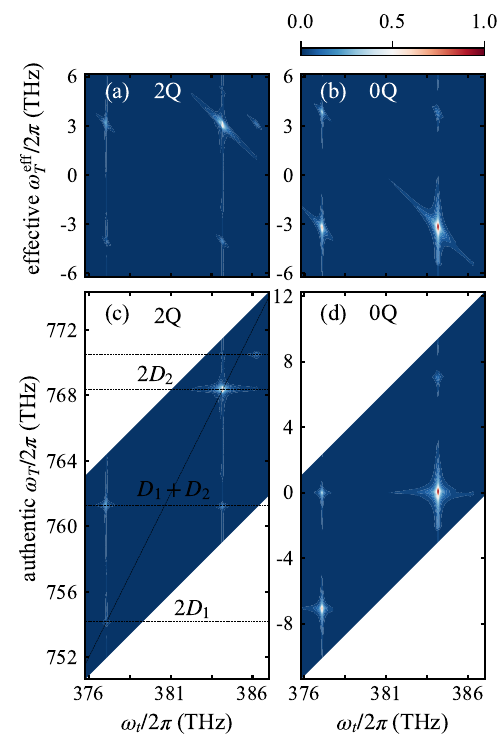}

\caption{2Q 2D absolute spectra and 0Q 2D absolute spectra of $^{87}\mathrm{Rb}$. (a) and (b) are 2Q and 0Q 2D spectra without frequency shifts retrieved, and their y-axes represent the effective frequency $\omega_{T}^{\mathrm{eff}}$. (c) and (d) are obtained by retrieving frequency shifts of (a) and (b), with their y-axes representing the authentic frequency $\omega_{T}$. In (c), the peaks on the highest horizontal dashed line correspond to the doubly excited level of an individual $^{87}\mathrm{Rb}$ atom, i.e., $5^{2}D$. The peaks on the other (lower) three horizontal dashed lines correspond to the doubly excited levels originating from dipole-dipole interactions.}
\label{spectra}
\end{figure}

By applying the separation process in Sec.~\ref{subsec:Enforcing-causality}, we obtain the separated 2Q and 0Q 2D spectra without frequency shifts retrieved, $S_{\mathrm{2Q}}^{\mathrm{sep}}(\tau,\omega_{T}^{\mathrm{eff}},\omega_{t})$ and $S_{\mathrm{0Q}}^{\mathrm{sep}}(\tau,\omega_{T}^{\mathrm{eff}},\omega_{t})$, as illustrated in Fig.~\ref{spectra}(a) and (b). Subsequently, the y-axes of these two spectra are added by $\omega_{t}+\omega_{\mathrm{rf}}$ and $\omega_{t}-\omega_{\mathrm{rf}}$, respectively, resulting in the authentic 2Q and 0Q 2D spectra, $S_{\mathrm{2Q}}^{\mathrm{rtrv}}(\tau,\omega_{T},\omega_{t})$ and $S_{\mathrm{0Q}}^{\mathrm{rtrv}}(\tau,\omega_{T},\omega_{t})$, as shown in Fig.~\ref{spectra}(c) and (d). Note that we additionally calibrate the axis of $\omega_{T}$ of the 2Q 2D spectrum by adding \added{$2\pi\times0.2\:\mathrm{THz}$} to have the \replaced{highest peak }{two highest peaks} settled at $\omega_{T}/2\pi=770.5\:\mathrm{THz}$~\citep{Gao2016} in Fig.~\ref{spectra}(c), where the highest horizontal dashed line denotes the frequency of $770.5\:\mathrm{THz}$. Such operation is performed to compensate for the shift in the calibration of nonlinearities (typically in the AOM) in the system.

After the frequency calibration, the 2Q and 0Q 2D spectra of $^{87}\mathrm{Rb}$ in Fig.~\ref{spectra}(c) and (d) show identical peak locations as those have been reported in other systems utilizing either BOXCARS or collinear geometries~\citep{Gao2016,Yan2022}. Specifically, in Fig.~\ref{spectra}(c), the \replaced{peak }{two peaks} aligned on the highest horizontal dashed line \added{at $(\omega_t / 2\pi = 386.3\:\mathrm{THz}, \omega_T / 2\pi = 770.5\:\mathrm{THz})$} corresponds to the doubly excited level of an individual $^{87}\mathrm{Rb}$ atom. The three lower horizontal dashed lines correspond to frequencies $754.2\:\mathrm{THz}$, $761.3\:\mathrm{THz}$, and $768.4\:\mathrm{THz}$, respectively, signifying the collective resonances originated from dipole-dipole interactions of $D_{1}+D_{1}$, $D_{1}+D_{2}$, and $D_{2}+D_{2}$ transitions of two $^{87}\mathrm{Rb}$ atoms. \added{Another peak corresponding to the doubly excited level of an individual $^{87}\mathrm{Rb}$ atom should be located at $(384.2\:\mathrm{THz}, 770.5\:\mathrm{THz})$ but is covered by the broadening of the adjacent stronger peak at $(384.2\:\mathrm{THz}, 768.4\:\mathrm{THz})$.} In Fig.~\ref{spectra}(d), two peaks on $\omega_{T}=0$ are contributed by population states during $T$, and the other two peaks on $\omega_{T}=\pm7.1\:\mathrm{THz}$ correspond to coherences between $D_{1}$ and $D_{2}$ transitions. The detailed double-sided Feynman diagrams of all the corresponding pathways are listed in the supplementary material. We note that we have firstly accomplished the simultaneous observation of the doubly excited level of an individual $^{87}\mathrm{Rb}$ atom and collective resonances arising from dipole-dipole interactions of both $D_{1}$ and $D_{2}$ transitions~\citep{Yan2022} within a shared 2Q 2D spectrum.

\section{Conclusion}

2D spectroscopy under pump-probe geometry is an approach alleviating the challenges of implementing 2D spectroscopy and has achieved considerable success in 1Q 2D spectroscopy since its invention. However, in the realm of the third-order 2Q studies using 2D spectroscopy, the pump-probe geometry has been limited due to its phase matching direction, while other geometries, such as the BOXCARS and collinear geometries, have historically taken precedence. This situation was only recently broken \added{with a permuted-pump-probe pulse sequence in 2D IR~\citep{Farrell2022} and 2D electronic~\citep{Armstrong2023} spectroscopies}. Our work extends the application of this method to 2Q 2DES, providing an alternative for \added{extracting} and investigating electronic 2Q coherence through 2D spectroscopy.

In our 2Q 2DES under pump-probe geometry, the probe pulse arrives firstly at the sample followed by the two pump pulses. In such a pulse sequence, not only the 2Q but also the 0Q coherences are detected due to their same phase matching direction. We have discussed in detail the effective oscillation frequencies of 2Q and 0Q coherences during $T$. We present that both 2Q and 0Q coherences may appear at either quadrants along the axis of $\omega_{T}^{\mathrm{eff}}$ on the 2D spectrum. Thus, we separate the 2Q and 0Q coherence into two distinct spectra by using the technique of causality enforcement and by applying the typical phase cycling scheme $(\phi_{1}=0,\phi_{1'}=0)$ and $(\phi_{1}=0,\phi_{1'}=\pi/2)$ of the two pump pulses. 2Q and 0Q spectra are then analyzed without disturbing each other.

\section*{Supplementary Material}
\added{Detailed double-sided Feynman diagrams are listed in the supplementary material.}

\begin{acknowledgments}
This work was supported by the National Natural Science Foundation of China (Grant Nos. U2230203, U2330401, 12088101).
\end{acknowledgments}

\section*{AUTHOR DECLARATIONS}

\subsection*{Conflict of Interest}

The authors have no conflicts to disclose.

\subsection*{Author Contributions}
\noindent{\bf Mao-Rui Cai}: Conceptualization (equal); Data Curation (lead); Formal Analysis (equal); Methodology (equal); Software (equal); Writing--original draft (lead); Writing--review \& editing (supporting). {\bf Xue Zhang}: Formal Analysis (equal); Methodology (equal); Writing--review \& editing (supporting). {\bf Zi-Qian Cheng}: Formal Analysis (equal); Writing--review \& editing (supporting). {\bf Teng-Fei Yan}: Software (equal); Writing--review \& editing (supporting). {\bf Hui Dong}: Conceptualization (equal); Formal Analysis (equal); Funding acquisition (lead); Methodology (equal); Supervision (lead); Project Administration (lead); Writing--review \& editing (lead).

\section*{DATA AVAILABILITY}
The data that support the findings of this study are available
from the corresponding author upon reasonable request.

\bibliographystyle{apsrev4-1}
\bibliography{Rb-2Q}

\begin{thebibliography}{52}%
\makeatletter
\providecommand \@ifxundefined [1]{%
 \@ifx{#1\undefined}
}%
\providecommand \@ifnum [1]{%
 \ifnum #1\expandafter \@firstoftwo
 \else \expandafter \@secondoftwo
 \fi
}%
\providecommand \@ifx [1]{%
 \ifx #1\expandafter \@firstoftwo
 \else \expandafter \@secondoftwo
 \fi
}%
\providecommand \natexlab [1]{#1}%
\providecommand \enquote  [1]{``#1''}%
\providecommand \bibnamefont  [1]{#1}%
\providecommand \bibfnamefont [1]{#1}%
\providecommand \citenamefont [1]{#1}%
\providecommand \href@noop [0]{\@secondoftwo}%
\providecommand \href [0]{\begingroup \@sanitize@url \@href}%
\providecommand \@href[1]{\@@startlink{#1}\@@href}%
\providecommand \@@href[1]{\endgroup#1\@@endlink}%
\providecommand \@sanitize@url [0]{\catcode `\\12\catcode `\$12\catcode
  `\&12\catcode `\#12\catcode `\^12\catcode `\_12\catcode `\%12\relax}%
\providecommand \@@startlink[1]{}%
\providecommand \@@endlink[0]{}%
\providecommand \url  [0]{\begingroup\@sanitize@url \@url }%
\providecommand \@url [1]{\endgroup\@href {#1}{\urlprefix }}%
\providecommand \urlprefix  [0]{URL }%
\providecommand \Eprint [0]{\href }%
\providecommand \doibase [0]{http://dx.doi.org/}%
\providecommand \selectlanguage [0]{\@gobble}%
\providecommand \bibinfo  [0]{\@secondoftwo}%
\providecommand \bibfield  [0]{\@secondoftwo}%
\providecommand \translation [1]{[#1]}%
\providecommand \BibitemOpen [0]{}%
\providecommand \bibitemStop [0]{}%
\providecommand \bibitemNoStop [0]{.\EOS\space}%
\providecommand \EOS [0]{\spacefactor3000\relax}%
\providecommand \BibitemShut  [1]{\csname bibitem#1\endcsname}%
\let\auto@bib@innerbib\@empty
\bibitem [{\citenamefont {Mukamel}(1995)}]{mukamel1995book}%
  \BibitemOpen
  \bibfield  {author} {\bibinfo {author} {\bibfnamefont {S.}~\bibnamefont
  {Mukamel}},\ }\href@noop {} {\emph {\bibinfo {title} {Principles of Nonlinear
  Optical Spectroscopy}}}\ (\bibinfo  {publisher} {Oxford University Press},\
  \bibinfo {year} {1995})\BibitemShut {NoStop}%
\bibitem [{\citenamefont {Cho}(2009)}]{cho2009book}%
  \BibitemOpen
  \bibfield  {author} {\bibinfo {author} {\bibfnamefont {M.}~\bibnamefont
  {Cho}},\ }\href@noop {} {\emph {\bibinfo {title} {Two-Dimensional Optical
  Spectroscopy}}}\ (\bibinfo  {publisher} {CRC Press},\ \bibinfo {year}
  {2009})\BibitemShut {NoStop}%
\bibitem [{\citenamefont {Hamm}\ and\ \citenamefont
  {Zanni}(2011)}]{hamm2011concepts}%
  \BibitemOpen
  \bibfield  {author} {\bibinfo {author} {\bibfnamefont {P.}~\bibnamefont
  {Hamm}}\ and\ \bibinfo {author} {\bibfnamefont {M.}~\bibnamefont {Zanni}},\
  }\href@noop {} {\emph {\bibinfo {title} {Concepts and Methods of 2D Infrared
  Spectroscopy}}}\ (\bibinfo  {publisher} {Cambridge University Press},\
  \bibinfo {year} {2011})\BibitemShut {NoStop}%
\bibitem [{\citenamefont {Jonas}(2003)}]{Jonas2003}%
  \BibitemOpen
  \bibfield  {author} {\bibinfo {author} {\bibfnamefont {D.~M.}\ \bibnamefont
  {Jonas}},\ }\href {\doibase 10.1146/annurev.physchem.54.011002.103907}
  {\bibfield  {journal} {\bibinfo  {journal} {Annual Review of Physical
  Chemistry}\ }\textbf {\bibinfo {volume} {54}},\ \bibinfo {pages} {425}
  (\bibinfo {year} {2003})}\BibitemShut {NoStop}%
\bibitem [{\citenamefont {Tian}\ \emph {et~al.}(2003)\citenamefont {Tian},
  \citenamefont {Keusters}, \citenamefont {Suzaki},\ and\ \citenamefont
  {Warren}}]{Tian2003}%
  \BibitemOpen
  \bibfield  {author} {\bibinfo {author} {\bibfnamefont {P.}~\bibnamefont
  {Tian}}, \bibinfo {author} {\bibfnamefont {D.}~\bibnamefont {Keusters}},
  \bibinfo {author} {\bibfnamefont {Y.}~\bibnamefont {Suzaki}}, \ and\ \bibinfo
  {author} {\bibfnamefont {W.~S.}\ \bibnamefont {Warren}},\ }\href {\doibase
  10.1126/science.1083433} {\bibfield  {journal} {\bibinfo  {journal}
  {Science}\ }\textbf {\bibinfo {volume} {300}},\ \bibinfo {pages} {1553}
  (\bibinfo {year} {2003})}\BibitemShut {NoStop}%
\bibitem [{\citenamefont {Dai}\ \emph {et~al.}(2012)\citenamefont {Dai},
  \citenamefont {Richter}, \citenamefont {Li}, \citenamefont {Bristow},
  \citenamefont {Falvo}, \citenamefont {Mukamel},\ and\ \citenamefont
  {Cundiff}}]{Dai2012}%
  \BibitemOpen
  \bibfield  {author} {\bibinfo {author} {\bibfnamefont {X.}~\bibnamefont
  {Dai}}, \bibinfo {author} {\bibfnamefont {M.}~\bibnamefont {Richter}},
  \bibinfo {author} {\bibfnamefont {H.}~\bibnamefont {Li}}, \bibinfo {author}
  {\bibfnamefont {A.~D.}\ \bibnamefont {Bristow}}, \bibinfo {author}
  {\bibfnamefont {C.}~\bibnamefont {Falvo}}, \bibinfo {author} {\bibfnamefont
  {S.}~\bibnamefont {Mukamel}}, \ and\ \bibinfo {author} {\bibfnamefont
  {S.~T.}\ \bibnamefont {Cundiff}},\ }\href {\doibase
  10.1103/PhysRevLett.108.193201} {\bibfield  {journal} {\bibinfo  {journal}
  {Phys. Rev. Lett.}\ }\textbf {\bibinfo {volume} {108}},\ \bibinfo {pages}
  {193201} (\bibinfo {year} {2012})}\BibitemShut {NoStop}%
\bibitem [{\citenamefont {Lomsadze}\ and\ \citenamefont
  {Cundiff}(2017)}]{Cundiff2017}%
  \BibitemOpen
  \bibfield  {author} {\bibinfo {author} {\bibfnamefont {B.}~\bibnamefont
  {Lomsadze}}\ and\ \bibinfo {author} {\bibfnamefont {S.~T.}\ \bibnamefont
  {Cundiff}},\ }\href {\doibase 10.1126/science.aao1090} {\bibfield  {journal}
  {\bibinfo  {journal} {Science}\ }\textbf {\bibinfo {volume} {357}},\ \bibinfo
  {pages} {1389} (\bibinfo {year} {2017})}\BibitemShut {NoStop}%
\bibitem [{\citenamefont {Lomsadze}\ and\ \citenamefont
  {Cundiff}(2018)}]{Lomsadze2018}%
  \BibitemOpen
  \bibfield  {author} {\bibinfo {author} {\bibfnamefont {B.}~\bibnamefont
  {Lomsadze}}\ and\ \bibinfo {author} {\bibfnamefont {S.~T.}\ \bibnamefont
  {Cundiff}},\ }\href {\doibase 10.1103/PhysRevLett.120.233401} {\bibfield
  {journal} {\bibinfo  {journal} {Phys. Rev. Lett.}\ }\textbf {\bibinfo
  {volume} {120}},\ \bibinfo {pages} {233401} (\bibinfo {year}
  {2018})}\BibitemShut {NoStop}%
\bibitem [{\citenamefont {Zhang}\ \emph {et~al.}(2021)\citenamefont {Zhang},
  \citenamefont {Shi}, \citenamefont {Li}, \citenamefont {Coy}, \citenamefont
  {Field},\ and\ \citenamefont {Nelson}}]{Nelson2021}%
  \BibitemOpen
  \bibfield  {author} {\bibinfo {author} {\bibfnamefont {Y.}~\bibnamefont
  {Zhang}}, \bibinfo {author} {\bibfnamefont {J.}~\bibnamefont {Shi}}, \bibinfo
  {author} {\bibfnamefont {X.}~\bibnamefont {Li}}, \bibinfo {author}
  {\bibfnamefont {S.~L.}\ \bibnamefont {Coy}}, \bibinfo {author} {\bibfnamefont
  {R.~W.}\ \bibnamefont {Field}}, \ and\ \bibinfo {author} {\bibfnamefont
  {K.~A.}\ \bibnamefont {Nelson}},\ }\href {\doibase 10.1073/pnas.2020941118}
  {\bibfield  {journal} {\bibinfo  {journal} {Proceedings of the National
  Academy of Sciences}\ }\textbf {\bibinfo {volume} {118}},\ \bibinfo {pages}
  {e2020941118} (\bibinfo {year} {2021})}\BibitemShut {NoStop}%
\bibitem [{\citenamefont {Schlau-Cohen}\ \emph {et~al.}(2011)\citenamefont
  {Schlau-Cohen}, \citenamefont {Ishizaki},\ and\ \citenamefont
  {Fleming}}]{Schlau-Cohen2011}%
  \BibitemOpen
  \bibfield  {author} {\bibinfo {author} {\bibfnamefont {G.~S.}\ \bibnamefont
  {Schlau-Cohen}}, \bibinfo {author} {\bibfnamefont {A.}~\bibnamefont
  {Ishizaki}}, \ and\ \bibinfo {author} {\bibfnamefont {G.~R.}\ \bibnamefont
  {Fleming}},\ }\href {\doibase https://doi.org/10.1016/j.chemphys.2011.04.025}
  {\bibfield  {journal} {\bibinfo  {journal} {Chemical Physics}\ }\textbf
  {\bibinfo {volume} {386}},\ \bibinfo {pages} {1} (\bibinfo {year}
  {2011})}\BibitemShut {NoStop}%
\bibitem [{\citenamefont {Elsaesser}(2009)}]{Thomas2009}%
  \BibitemOpen
  \bibfield  {author} {\bibinfo {author} {\bibfnamefont {T.}~\bibnamefont
  {Elsaesser}},\ }\href {\doibase 10.1021/ar900006u} {\bibfield  {journal}
  {\bibinfo  {journal} {Accounts of Chemical Research}\ }\textbf {\bibinfo
  {volume} {42}},\ \bibinfo {pages} {1220} (\bibinfo {year}
  {2009})}\BibitemShut {NoStop}%
\bibitem [{\citenamefont {Asplund}\ \emph {et~al.}(2000)\citenamefont
  {Asplund}, \citenamefont {Zanni},\ and\ \citenamefont
  {Hochstrasser}}]{Hochstrasser2000}%
  \BibitemOpen
  \bibfield  {author} {\bibinfo {author} {\bibfnamefont {M.~C.}\ \bibnamefont
  {Asplund}}, \bibinfo {author} {\bibfnamefont {M.~T.}\ \bibnamefont {Zanni}},
  \ and\ \bibinfo {author} {\bibfnamefont {R.~M.}\ \bibnamefont
  {Hochstrasser}},\ }\href {\doibase 10.1073/pnas.140227997} {\bibfield
  {journal} {\bibinfo  {journal} {Proceedings of the National Academy of
  Sciences}\ }\textbf {\bibinfo {volume} {97}},\ \bibinfo {pages} {8219}
  (\bibinfo {year} {2000})}\BibitemShut {NoStop}%
\bibitem [{\citenamefont {Demird\"oven}\ \emph {et~al.}(2002)\citenamefont
  {Demird\"oven}, \citenamefont {Khalil},\ and\ \citenamefont
  {Tokmakoff}}]{Tokmakoff2002}%
  \BibitemOpen
  \bibfield  {author} {\bibinfo {author} {\bibfnamefont {N.}~\bibnamefont
  {Demird\"oven}}, \bibinfo {author} {\bibfnamefont {M.}~\bibnamefont
  {Khalil}}, \ and\ \bibinfo {author} {\bibfnamefont {A.}~\bibnamefont
  {Tokmakoff}},\ }\href {\doibase 10.1103/PhysRevLett.89.237401} {\bibfield
  {journal} {\bibinfo  {journal} {Phys. Rev. Lett.}\ }\textbf {\bibinfo
  {volume} {89}},\ \bibinfo {pages} {237401} (\bibinfo {year}
  {2002})}\BibitemShut {NoStop}%
\bibitem [{\citenamefont {Middleton}\ \emph {et~al.}(2010)\citenamefont
  {Middleton}, \citenamefont {Woys}, \citenamefont {Mukherjee},\ and\
  \citenamefont {Zanni}}]{Chris2010}%
  \BibitemOpen
  \bibfield  {author} {\bibinfo {author} {\bibfnamefont {C.~T.}\ \bibnamefont
  {Middleton}}, \bibinfo {author} {\bibfnamefont {A.~M.}\ \bibnamefont {Woys}},
  \bibinfo {author} {\bibfnamefont {S.~S.}\ \bibnamefont {Mukherjee}}, \ and\
  \bibinfo {author} {\bibfnamefont {M.~T.}\ \bibnamefont {Zanni}},\ }\href
  {\doibase https://doi.org/10.1016/j.ymeth.2010.05.002} {\bibfield  {journal}
  {\bibinfo  {journal} {Methods}\ }\textbf {\bibinfo {volume} {52}},\ \bibinfo
  {pages} {12} (\bibinfo {year} {2010})}\BibitemShut {NoStop}%
\bibitem [{\citenamefont {Middleton}\ \emph {et~al.}(2012)\citenamefont
  {Middleton}, \citenamefont {Marek}, \citenamefont {Cao}, \citenamefont
  {Chiu}, \citenamefont {Singh}, \citenamefont {Woys}, \citenamefont
  {de~Pablo}, \citenamefont {Raleigh},\ and\ \citenamefont
  {Zanni}}]{Chris2012}%
  \BibitemOpen
  \bibfield  {author} {\bibinfo {author} {\bibfnamefont {C.~T.}\ \bibnamefont
  {Middleton}}, \bibinfo {author} {\bibfnamefont {P.}~\bibnamefont {Marek}},
  \bibinfo {author} {\bibfnamefont {P.}~\bibnamefont {Cao}}, \bibinfo {author}
  {\bibfnamefont {C.-c.}\ \bibnamefont {Chiu}}, \bibinfo {author}
  {\bibfnamefont {S.}~\bibnamefont {Singh}}, \bibinfo {author} {\bibfnamefont
  {A.~M.}\ \bibnamefont {Woys}}, \bibinfo {author} {\bibfnamefont {J.~J.}\
  \bibnamefont {de~Pablo}}, \bibinfo {author} {\bibfnamefont {D.~P.}\
  \bibnamefont {Raleigh}}, \ and\ \bibinfo {author} {\bibfnamefont {M.~T.}\
  \bibnamefont {Zanni}},\ }\href {\doibase 10.1038/nchem.1293} {\bibfield
  {journal} {\bibinfo  {journal} {Nature Chemistry}\ }\textbf {\bibinfo
  {volume} {4}},\ \bibinfo {pages} {355} (\bibinfo {year} {2012})}\BibitemShut
  {NoStop}%
\bibitem [{\citenamefont {Cowan}\ \emph {et~al.}(2004)\citenamefont {Cowan},
  \citenamefont {Ogilvie},\ and\ \citenamefont {Miller}}]{Miller2004}%
  \BibitemOpen
  \bibfield  {author} {\bibinfo {author} {\bibfnamefont {M.}~\bibnamefont
  {Cowan}}, \bibinfo {author} {\bibfnamefont {J.}~\bibnamefont {Ogilvie}}, \
  and\ \bibinfo {author} {\bibfnamefont {R.}~\bibnamefont {Miller}},\ }\href
  {\doibase https://doi.org/10.1016/j.cplett.2004.01.027} {\bibfield  {journal}
  {\bibinfo  {journal} {Chemical Physics Letters}\ }\textbf {\bibinfo {volume}
  {386}},\ \bibinfo {pages} {184} (\bibinfo {year} {2004})}\BibitemShut
  {NoStop}%
\bibitem [{\citenamefont {Brixner}\ \emph {et~al.}(2005)\citenamefont
  {Brixner}, \citenamefont {Stenger}, \citenamefont {Vaswani}, \citenamefont
  {Cho}, \citenamefont {Blankenship},\ and\ \citenamefont
  {Fleming}}]{Fleming2005}%
  \BibitemOpen
  \bibfield  {author} {\bibinfo {author} {\bibfnamefont {T.}~\bibnamefont
  {Brixner}}, \bibinfo {author} {\bibfnamefont {J.}~\bibnamefont {Stenger}},
  \bibinfo {author} {\bibfnamefont {H.~M.}\ \bibnamefont {Vaswani}}, \bibinfo
  {author} {\bibfnamefont {M.}~\bibnamefont {Cho}}, \bibinfo {author}
  {\bibfnamefont {R.~E.}\ \bibnamefont {Blankenship}}, \ and\ \bibinfo {author}
  {\bibfnamefont {G.~R.}\ \bibnamefont {Fleming}},\ }\href {\doibase
  10.1038/nature03429} {\bibfield  {journal} {\bibinfo  {journal} {Nature}\
  }\textbf {\bibinfo {volume} {434}},\ \bibinfo {pages} {625} (\bibinfo {year}
  {2005})}\BibitemShut {NoStop}%
\bibitem [{\citenamefont {Lewis}\ \emph {et~al.}(2016)\citenamefont {Lewis},
  \citenamefont {Gruenke}, \citenamefont {Oliver}, \citenamefont {Ballottari},
  \citenamefont {Bassi},\ and\ \citenamefont {Fleming}}]{Fleming2016}%
  \BibitemOpen
  \bibfield  {author} {\bibinfo {author} {\bibfnamefont {N.~H.~C.}\
  \bibnamefont {Lewis}}, \bibinfo {author} {\bibfnamefont {N.~L.}\ \bibnamefont
  {Gruenke}}, \bibinfo {author} {\bibfnamefont {T.~A.~A.}\ \bibnamefont
  {Oliver}}, \bibinfo {author} {\bibfnamefont {M.}~\bibnamefont {Ballottari}},
  \bibinfo {author} {\bibfnamefont {R.}~\bibnamefont {Bassi}}, \ and\ \bibinfo
  {author} {\bibfnamefont {G.~R.}\ \bibnamefont {Fleming}},\ }\href {\doibase
  10.1021/acs.jpclett.6b02280} {\bibfield  {journal} {\bibinfo  {journal} {The
  Journal of Physical Chemistry Letters}\ }\textbf {\bibinfo {volume} {7}},\
  \bibinfo {pages} {4197} (\bibinfo {year} {2016})}\BibitemShut {NoStop}%
\bibitem [{\citenamefont {hung Tseng}\ \emph {et~al.}(2009)\citenamefont {hung
  Tseng}, \citenamefont {Matsika},\ and\ \citenamefont {Weinacht}}]{Tseng2009}%
  \BibitemOpen
  \bibfield  {author} {\bibinfo {author} {\bibfnamefont {C.}~\bibnamefont {hung
  Tseng}}, \bibinfo {author} {\bibfnamefont {S.}~\bibnamefont {Matsika}}, \
  and\ \bibinfo {author} {\bibfnamefont {T.~C.}\ \bibnamefont {Weinacht}},\
  }\href {\doibase 10.1364/OE.17.018788} {\bibfield  {journal} {\bibinfo
  {journal} {Opt. Express}\ }\textbf {\bibinfo {volume} {17}},\ \bibinfo
  {pages} {18788} (\bibinfo {year} {2009})}\BibitemShut {NoStop}%
\bibitem [{\citenamefont {West}\ and\ \citenamefont {Moran}(2012)}]{Moran2012}%
  \BibitemOpen
  \bibfield  {author} {\bibinfo {author} {\bibfnamefont {B.~A.}\ \bibnamefont
  {West}}\ and\ \bibinfo {author} {\bibfnamefont {A.~M.}\ \bibnamefont
  {Moran}},\ }\href {\doibase 10.1021/jz301048n} {\bibfield  {journal}
  {\bibinfo  {journal} {The Journal of Physical Chemistry Letters}\ }\textbf
  {\bibinfo {volume} {3}},\ \bibinfo {pages} {2575} (\bibinfo {year}
  {2012})}\BibitemShut {NoStop}%
\bibitem [{\citenamefont {Lu}\ \emph {et~al.}(2016)\citenamefont {Lu},
  \citenamefont {Zhang}, \citenamefont {Hwang}, \citenamefont {Ofori-Okai},
  \citenamefont {Fleischer},\ and\ \citenamefont {Nelson}}]{Lu2016}%
  \BibitemOpen
  \bibfield  {author} {\bibinfo {author} {\bibfnamefont {J.}~\bibnamefont
  {Lu}}, \bibinfo {author} {\bibfnamefont {Y.}~\bibnamefont {Zhang}}, \bibinfo
  {author} {\bibfnamefont {H.~Y.}\ \bibnamefont {Hwang}}, \bibinfo {author}
  {\bibfnamefont {B.~K.}\ \bibnamefont {Ofori-Okai}}, \bibinfo {author}
  {\bibfnamefont {S.}~\bibnamefont {Fleischer}}, \ and\ \bibinfo {author}
  {\bibfnamefont {K.~A.}\ \bibnamefont {Nelson}},\ }\href {\doibase
  10.1073/pnas.1609558113} {\bibfield  {journal} {\bibinfo  {journal}
  {Proceedings of the National Academy of Sciences}\ }\textbf {\bibinfo
  {volume} {113}},\ \bibinfo {pages} {11800} (\bibinfo {year}
  {2016})}\BibitemShut {NoStop}%
\bibitem [{\citenamefont {Reimann}\ \emph {et~al.}(2021)\citenamefont
  {Reimann}, \citenamefont {Woerner},\ and\ \citenamefont
  {Elsaesser}}]{Reimann2021}%
  \BibitemOpen
  \bibfield  {author} {\bibinfo {author} {\bibfnamefont {K.}~\bibnamefont
  {Reimann}}, \bibinfo {author} {\bibfnamefont {M.}~\bibnamefont {Woerner}}, \
  and\ \bibinfo {author} {\bibfnamefont {T.}~\bibnamefont {Elsaesser}},\ }\href
  {https://doi.org/10.1063/5.0046664} {\bibfield  {journal} {\bibinfo
  {journal} {The Journal of Chemical Physics}\ }\textbf {\bibinfo {volume}
  {154}} (\bibinfo {year} {2021})}\BibitemShut {NoStop}%
\bibitem [{\citenamefont {Oliver}\ \emph {et~al.}(2014)\citenamefont {Oliver},
  \citenamefont {Lewis},\ and\ \citenamefont {Fleming}}]{Thomas2014}%
  \BibitemOpen
  \bibfield  {author} {\bibinfo {author} {\bibfnamefont {T.~A.~A.}\
  \bibnamefont {Oliver}}, \bibinfo {author} {\bibfnamefont {N.~H.~C.}\
  \bibnamefont {Lewis}}, \ and\ \bibinfo {author} {\bibfnamefont {G.~R.}\
  \bibnamefont {Fleming}},\ }\href {\doibase 10.1073/pnas.1409207111}
  {\bibfield  {journal} {\bibinfo  {journal} {Proceedings of the National
  Academy of Sciences}\ }\textbf {\bibinfo {volume} {111}},\ \bibinfo {pages}
  {10061} (\bibinfo {year} {2014})}\BibitemShut {NoStop}%
\bibitem [{\citenamefont {Dong}\ \emph {et~al.}(2015)\citenamefont {Dong},
  \citenamefont {Lewis}, \citenamefont {Oliver},\ and\ \citenamefont
  {Fleming}}]{Dong2015}%
  \BibitemOpen
  \bibfield  {author} {\bibinfo {author} {\bibfnamefont {H.}~\bibnamefont
  {Dong}}, \bibinfo {author} {\bibfnamefont {N.~H.~C.}\ \bibnamefont {Lewis}},
  \bibinfo {author} {\bibfnamefont {T.~A.~A.}\ \bibnamefont {Oliver}}, \ and\
  \bibinfo {author} {\bibfnamefont {G.~R.}\ \bibnamefont {Fleming}},\ }\href
  {\doibase 10.1063/1.4919684} {\bibfield  {journal} {\bibinfo  {journal} {The
  Journal of Chemical Physics}\ }\textbf {\bibinfo {volume} {142}},\ \bibinfo
  {pages} {174201} (\bibinfo {year} {2015})}\BibitemShut {NoStop}%
\bibitem [{\citenamefont {Lewis}\ \emph {et~al.}(2015)\citenamefont {Lewis},
  \citenamefont {Dong}, \citenamefont {Oliver},\ and\ \citenamefont
  {Fleming}}]{Lewis2015}%
  \BibitemOpen
  \bibfield  {author} {\bibinfo {author} {\bibfnamefont {N.~H.~C.}\
  \bibnamefont {Lewis}}, \bibinfo {author} {\bibfnamefont {H.}~\bibnamefont
  {Dong}}, \bibinfo {author} {\bibfnamefont {T.~A.~A.}\ \bibnamefont {Oliver}},
  \ and\ \bibinfo {author} {\bibfnamefont {G.~R.}\ \bibnamefont {Fleming}},\
  }\href {\doibase 10.1063/1.4919686} {\bibfield  {journal} {\bibinfo
  {journal} {The Journal of Chemical Physics}\ }\textbf {\bibinfo {volume}
  {142}},\ \bibinfo {pages} {174202} (\bibinfo {year} {2015})}\BibitemShut
  {NoStop}%
\bibitem [{\citenamefont {Courtney}\ \emph {et~al.}(2015)\citenamefont
  {Courtney}, \citenamefont {Fox}, \citenamefont {Slenkamp},\ and\
  \citenamefont {Khalil}}]{Courtney2015}%
  \BibitemOpen
  \bibfield  {author} {\bibinfo {author} {\bibfnamefont {T.~L.}\ \bibnamefont
  {Courtney}}, \bibinfo {author} {\bibfnamefont {Z.~W.}\ \bibnamefont {Fox}},
  \bibinfo {author} {\bibfnamefont {K.~M.}\ \bibnamefont {Slenkamp}}, \ and\
  \bibinfo {author} {\bibfnamefont {M.}~\bibnamefont {Khalil}},\ }\href
  {\doibase 10.1063/1.4932983} {\bibfield  {journal} {\bibinfo  {journal} {The
  Journal of Chemical Physics}\ }\textbf {\bibinfo {volume} {143}},\ \bibinfo
  {pages} {154201} (\bibinfo {year} {2015})}\BibitemShut {NoStop}%
\bibitem [{\citenamefont {Cundiff}\ and\ \citenamefont
  {Mukamel}(2013)}]{CundiffMukamel2013}%
  \BibitemOpen
  \bibfield  {author} {\bibinfo {author} {\bibfnamefont {S.~T.}\ \bibnamefont
  {Cundiff}}\ and\ \bibinfo {author} {\bibfnamefont {S.}~\bibnamefont
  {Mukamel}},\ }\href {\doibase 10.1063/PT.3.2047} {\bibfield  {journal}
  {\bibinfo  {journal} {Physics Today}\ }\textbf {\bibinfo {volume} {66}},\
  \bibinfo {pages} {44} (\bibinfo {year} {2013})}\BibitemShut {NoStop}%
\bibitem [{\citenamefont {Kim}\ \emph {et~al.}(2009)\citenamefont {Kim},
  \citenamefont {Mukamel},\ and\ \citenamefont {Scholes}}]{Kim2009}%
  \BibitemOpen
  \bibfield  {author} {\bibinfo {author} {\bibfnamefont {J.}~\bibnamefont
  {Kim}}, \bibinfo {author} {\bibfnamefont {S.}~\bibnamefont {Mukamel}}, \ and\
  \bibinfo {author} {\bibfnamefont {G.~D.}\ \bibnamefont {Scholes}},\ }\href
  {\doibase 10.1021/ar9000795} {\bibfield  {journal} {\bibinfo  {journal}
  {Accounts of Chemical Research}\ }\textbf {\bibinfo {volume} {42}},\ \bibinfo
  {pages} {1375} (\bibinfo {year} {2009})}\BibitemShut {NoStop}%
\bibitem [{\citenamefont {Brixner}\ \emph {et~al.}(2004)\citenamefont
  {Brixner}, \citenamefont {Mančal}, \citenamefont {Stiopkin},\ and\
  \citenamefont {Fleming}}]{Tobias2004}%
  \BibitemOpen
  \bibfield  {author} {\bibinfo {author} {\bibfnamefont {T.}~\bibnamefont
  {Brixner}}, \bibinfo {author} {\bibfnamefont {T.}~\bibnamefont {Mančal}},
  \bibinfo {author} {\bibfnamefont {I.~V.}\ \bibnamefont {Stiopkin}}, \ and\
  \bibinfo {author} {\bibfnamefont {G.~R.}\ \bibnamefont {Fleming}},\ }\href
  {\doibase 10.1063/1.1776112} {\bibfield  {journal} {\bibinfo  {journal} {The
  Journal of Chemical Physics}\ }\textbf {\bibinfo {volume} {121}},\ \bibinfo
  {pages} {4221} (\bibinfo {year} {2004})}\BibitemShut {NoStop}%
\bibitem [{\citenamefont {Stone}\ \emph {et~al.}(2009)\citenamefont {Stone},
  \citenamefont {Gundogdu}, \citenamefont {Turner}, \citenamefont {Li},
  \citenamefont {Cundiff},\ and\ \citenamefont {Nelson}}]{Stone2009}%
  \BibitemOpen
  \bibfield  {author} {\bibinfo {author} {\bibfnamefont {K.~W.}\ \bibnamefont
  {Stone}}, \bibinfo {author} {\bibfnamefont {K.}~\bibnamefont {Gundogdu}},
  \bibinfo {author} {\bibfnamefont {D.~B.}\ \bibnamefont {Turner}}, \bibinfo
  {author} {\bibfnamefont {X.}~\bibnamefont {Li}}, \bibinfo {author}
  {\bibfnamefont {S.~T.}\ \bibnamefont {Cundiff}}, \ and\ \bibinfo {author}
  {\bibfnamefont {K.~A.}\ \bibnamefont {Nelson}},\ }\href {\doibase
  10.1126/science.1170274} {\bibfield  {journal} {\bibinfo  {journal}
  {Science}\ }\textbf {\bibinfo {volume} {324}},\ \bibinfo {pages} {1169}
  (\bibinfo {year} {2009})}\BibitemShut {NoStop}%
\bibitem [{\citenamefont {Grumstrup}\ \emph {et~al.}(2007)\citenamefont
  {Grumstrup}, \citenamefont {Shim}, \citenamefont {Montgomery}, \citenamefont
  {Damrauer},\ and\ \citenamefont {Zanni}}]{Grumstrup2007}%
  \BibitemOpen
  \bibfield  {author} {\bibinfo {author} {\bibfnamefont {E.~M.}\ \bibnamefont
  {Grumstrup}}, \bibinfo {author} {\bibfnamefont {S.-H.}\ \bibnamefont {Shim}},
  \bibinfo {author} {\bibfnamefont {M.~A.}\ \bibnamefont {Montgomery}},
  \bibinfo {author} {\bibfnamefont {N.~H.}\ \bibnamefont {Damrauer}}, \ and\
  \bibinfo {author} {\bibfnamefont {M.~T.}\ \bibnamefont {Zanni}},\ }\href
  {\doibase 10.1364/OE.15.016681} {\bibfield  {journal} {\bibinfo  {journal}
  {Opt. Express}\ }\textbf {\bibinfo {volume} {15}},\ \bibinfo {pages} {16681}
  (\bibinfo {year} {2007})}\BibitemShut {NoStop}%
\bibitem [{\citenamefont {Myers}\ \emph {et~al.}(2008)\citenamefont {Myers},
  \citenamefont {Lewis}, \citenamefont {Tekavec},\ and\ \citenamefont
  {Ogilvie}}]{Myers2008}%
  \BibitemOpen
  \bibfield  {author} {\bibinfo {author} {\bibfnamefont {J.~A.}\ \bibnamefont
  {Myers}}, \bibinfo {author} {\bibfnamefont {K.~L.~M.}\ \bibnamefont {Lewis}},
  \bibinfo {author} {\bibfnamefont {P.~F.}\ \bibnamefont {Tekavec}}, \ and\
  \bibinfo {author} {\bibfnamefont {J.~P.}\ \bibnamefont {Ogilvie}},\ }\href
  {\doibase 10.1364/OE.16.017420} {\bibfield  {journal} {\bibinfo  {journal}
  {Opt. Express}\ }\textbf {\bibinfo {volume} {16}},\ \bibinfo {pages} {17420}
  (\bibinfo {year} {2008})}\BibitemShut {NoStop}%
\bibitem [{\citenamefont {Brida}\ \emph {et~al.}(2012)\citenamefont {Brida},
  \citenamefont {Manzoni},\ and\ \citenamefont {Cerullo}}]{Brida2012}%
  \BibitemOpen
  \bibfield  {author} {\bibinfo {author} {\bibfnamefont {D.}~\bibnamefont
  {Brida}}, \bibinfo {author} {\bibfnamefont {C.}~\bibnamefont {Manzoni}}, \
  and\ \bibinfo {author} {\bibfnamefont {G.}~\bibnamefont {Cerullo}},\ }\href
  {\doibase 10.1364/OL.37.003027} {\bibfield  {journal} {\bibinfo  {journal}
  {Opt. Lett.}\ }\textbf {\bibinfo {volume} {37}},\ \bibinfo {pages} {3027}
  (\bibinfo {year} {2012})}\BibitemShut {NoStop}%
\bibitem [{\citenamefont {Rock}\ \emph {et~al.}(2013)\citenamefont {Rock},
  \citenamefont {Li}, \citenamefont {Pagano},\ and\ \citenamefont
  {Cheatum}}]{Rock2013}%
  \BibitemOpen
  \bibfield  {author} {\bibinfo {author} {\bibfnamefont {W.}~\bibnamefont
  {Rock}}, \bibinfo {author} {\bibfnamefont {Y.-L.}\ \bibnamefont {Li}},
  \bibinfo {author} {\bibfnamefont {P.}~\bibnamefont {Pagano}}, \ and\ \bibinfo
  {author} {\bibfnamefont {C.~M.}\ \bibnamefont {Cheatum}},\ }\href {\doibase
  10.1021/jp312817t} {\bibfield  {journal} {\bibinfo  {journal} {The Journal of
  Physical Chemistry A}\ }\textbf {\bibinfo {volume} {117}},\ \bibinfo {pages}
  {6073} (\bibinfo {year} {2013})}\BibitemShut {NoStop}%
\bibitem [{\citenamefont {Zhang}\ \emph {et~al.}(1999)\citenamefont {Zhang},
  \citenamefont {Chernyak},\ and\ \citenamefont {Mukamel}}]{Zhang1999}%
  \BibitemOpen
  \bibfield  {author} {\bibinfo {author} {\bibfnamefont {W.~M.}\ \bibnamefont
  {Zhang}}, \bibinfo {author} {\bibfnamefont {V.}~\bibnamefont {Chernyak}}, \
  and\ \bibinfo {author} {\bibfnamefont {S.}~\bibnamefont {Mukamel}},\ }\href
  {\doibase 10.1063/1.478400} {\bibfield  {journal} {\bibinfo  {journal} {The
  Journal of Chemical Physics}\ }\textbf {\bibinfo {volume} {110}},\ \bibinfo
  {pages} {5011} (\bibinfo {year} {1999})}\BibitemShut {NoStop}%
\bibitem [{\citenamefont {Fulmer}\ \emph {et~al.}(2004)\citenamefont {Fulmer},
  \citenamefont {Mukherjee}, \citenamefont {Krummel},\ and\ \citenamefont
  {Zanni}}]{Fulmer2004}%
  \BibitemOpen
  \bibfield  {author} {\bibinfo {author} {\bibfnamefont {E.~C.}\ \bibnamefont
  {Fulmer}}, \bibinfo {author} {\bibfnamefont {P.}~\bibnamefont {Mukherjee}},
  \bibinfo {author} {\bibfnamefont {A.~T.}\ \bibnamefont {Krummel}}, \ and\
  \bibinfo {author} {\bibfnamefont {M.~T.}\ \bibnamefont {Zanni}},\ }\href
  {\doibase 10.1063/1.1649725} {\bibfield  {journal} {\bibinfo  {journal} {The
  Journal of Chemical Physics}\ }\textbf {\bibinfo {volume} {120}},\ \bibinfo
  {pages} {8067} (\bibinfo {year} {2004})}\BibitemShut {NoStop}%
\bibitem [{\citenamefont {Gao}\ \emph {et~al.}(2016)\citenamefont {Gao},
  \citenamefont {Cundiff},\ and\ \citenamefont {Li}}]{Gao2016}%
  \BibitemOpen
  \bibfield  {author} {\bibinfo {author} {\bibfnamefont {F.}~\bibnamefont
  {Gao}}, \bibinfo {author} {\bibfnamefont {S.~T.}\ \bibnamefont {Cundiff}}, \
  and\ \bibinfo {author} {\bibfnamefont {H.}~\bibnamefont {Li}},\ }\href
  {\doibase 10.1364/OL.41.002954} {\bibfield  {journal} {\bibinfo  {journal}
  {Opt. Lett.}\ }\textbf {\bibinfo {volume} {41}},\ \bibinfo {pages} {2954}
  (\bibinfo {year} {2016})}\BibitemShut {NoStop}%
\bibitem [{\citenamefont {Yu}\ \emph {et~al.}(2022)\citenamefont {Yu},
  \citenamefont {Geng}, \citenamefont {Liang}, \citenamefont {Li},\ and\
  \citenamefont {Liu}}]{Yu2022}%
  \BibitemOpen
  \bibfield  {author} {\bibinfo {author} {\bibfnamefont {S.}~\bibnamefont
  {Yu}}, \bibinfo {author} {\bibfnamefont {Y.}~\bibnamefont {Geng}}, \bibinfo
  {author} {\bibfnamefont {D.}~\bibnamefont {Liang}}, \bibinfo {author}
  {\bibfnamefont {H.}~\bibnamefont {Li}}, \ and\ \bibinfo {author}
  {\bibfnamefont {X.}~\bibnamefont {Liu}},\ }\href {\doibase 10.1364/OL.449365}
  {\bibfield  {journal} {\bibinfo  {journal} {Opt. Lett.}\ }\textbf {\bibinfo
  {volume} {47}},\ \bibinfo {pages} {997} (\bibinfo {year} {2022})}\BibitemShut
  {NoStop}%
\bibitem [{\citenamefont {Turner}\ and\ \citenamefont
  {Nelson}(2010)}]{Turner2010}%
  \BibitemOpen
  \bibfield  {author} {\bibinfo {author} {\bibfnamefont {D.~B.}\ \bibnamefont
  {Turner}}\ and\ \bibinfo {author} {\bibfnamefont {K.~A.}\ \bibnamefont
  {Nelson}},\ }\href {\doibase 10.1038/nature09286} {\bibfield  {journal}
  {\bibinfo  {journal} {Nature}\ }\textbf {\bibinfo {volume} {466}},\ \bibinfo
  {pages} {1089} (\bibinfo {year} {2010})}\BibitemShut {NoStop}%
\bibitem [{\citenamefont {Yu}\ \emph {et~al.}(2019)\citenamefont {Yu},
  \citenamefont {Titze}, \citenamefont {Zhu}, \citenamefont {Liu},\ and\
  \citenamefont {Li}}]{Yu2019}%
  \BibitemOpen
  \bibfield  {author} {\bibinfo {author} {\bibfnamefont {S.}~\bibnamefont
  {Yu}}, \bibinfo {author} {\bibfnamefont {M.}~\bibnamefont {Titze}}, \bibinfo
  {author} {\bibfnamefont {Y.}~\bibnamefont {Zhu}}, \bibinfo {author}
  {\bibfnamefont {X.}~\bibnamefont {Liu}}, \ and\ \bibinfo {author}
  {\bibfnamefont {H.}~\bibnamefont {Li}},\ }\href {\doibase
  10.1364/OL.44.002795} {\bibfield  {journal} {\bibinfo  {journal} {Opt.
  Lett.}\ }\textbf {\bibinfo {volume} {44}},\ \bibinfo {pages} {2795} (\bibinfo
  {year} {2019})}\BibitemShut {NoStop}%
\bibitem [{\citenamefont {Brosseau}\ \emph {et~al.}(2020)\citenamefont
  {Brosseau}, \citenamefont {Palato}, \citenamefont {Seiler}, \citenamefont
  {Baker},\ and\ \citenamefont {Kambhampati}}]{Patrick2020}%
  \BibitemOpen
  \bibfield  {author} {\bibinfo {author} {\bibfnamefont {P.}~\bibnamefont
  {Brosseau}}, \bibinfo {author} {\bibfnamefont {S.}~\bibnamefont {Palato}},
  \bibinfo {author} {\bibfnamefont {H.}~\bibnamefont {Seiler}}, \bibinfo
  {author} {\bibfnamefont {H.}~\bibnamefont {Baker}}, \ and\ \bibinfo {author}
  {\bibfnamefont {P.}~\bibnamefont {Kambhampati}},\ }\href {\doibase
  10.1063/5.0021381} {\bibfield  {journal} {\bibinfo  {journal} {The Journal of
  Chemical Physics}\ }\textbf {\bibinfo {volume} {153}},\ \bibinfo {pages}
  {234703} (\bibinfo {year} {2020})}\BibitemShut {NoStop}%
\bibitem [{\citenamefont {Farrell}\ and\ \citenamefont
  {Zanni}(2022)}]{Farrell2022}%
  \BibitemOpen
  \bibfield  {author} {\bibinfo {author} {\bibfnamefont {K.~M.}\ \bibnamefont
  {Farrell}}\ and\ \bibinfo {author} {\bibfnamefont {M.~T.}\ \bibnamefont
  {Zanni}},\ }\href {\doibase 10.1063/5.0097019} {\bibfield  {journal}
  {\bibinfo  {journal} {The Journal of Chemical Physics}\ }\textbf {\bibinfo
  {volume} {157}},\ \bibinfo {pages} {014203} (\bibinfo {year}
  {2022})}\BibitemShut {NoStop}%
\bibitem [{\citenamefont {Armstrong}\ \emph {et~al.}(2023)\citenamefont
  {Armstrong}, \citenamefont {Forlano}, \citenamefont {Roy}, \citenamefont
  {Bohlmann~Kunz}, \citenamefont {Farrell}, \citenamefont {Pan}, \citenamefont
  {Wright}, \citenamefont {Jin},\ and\ \citenamefont {Zanni}}]{Armstrong2023}%
  \BibitemOpen
  \bibfield  {author} {\bibinfo {author} {\bibfnamefont {Z.~T.}\ \bibnamefont
  {Armstrong}}, \bibinfo {author} {\bibfnamefont {K.~M.}\ \bibnamefont
  {Forlano}}, \bibinfo {author} {\bibfnamefont {C.~R.}\ \bibnamefont {Roy}},
  \bibinfo {author} {\bibfnamefont {M.}~\bibnamefont {Bohlmann~Kunz}}, \bibinfo
  {author} {\bibfnamefont {K.}~\bibnamefont {Farrell}}, \bibinfo {author}
  {\bibfnamefont {D.}~\bibnamefont {Pan}}, \bibinfo {author} {\bibfnamefont
  {J.~C.}\ \bibnamefont {Wright}}, \bibinfo {author} {\bibfnamefont
  {S.}~\bibnamefont {Jin}}, \ and\ \bibinfo {author} {\bibfnamefont {M.~T.}\
  \bibnamefont {Zanni}},\ }\href {\doibase 10.1021/jacs.3c05533} {\bibfield
  {journal} {\bibinfo  {journal} {Journal of the American Chemical Society}\
  }\textbf {\bibinfo {volume} {145}},\ \bibinfo {pages} {18568} (\bibinfo
  {year} {2023})}\BibitemShut {NoStop}%
\bibitem [{\citenamefont {Shim}\ and\ \citenamefont {Zanni}(2009)}]{Shim2009}%
  \BibitemOpen
  \bibfield  {author} {\bibinfo {author} {\bibfnamefont {S.-H.}\ \bibnamefont
  {Shim}}\ and\ \bibinfo {author} {\bibfnamefont {M.~T.}\ \bibnamefont
  {Zanni}},\ }\href {\doibase 10.1039/B813817F} {\bibfield  {journal} {\bibinfo
   {journal} {Phys. Chem. Chem. Phys.}\ }\textbf {\bibinfo {volume} {11}},\
  \bibinfo {pages} {748} (\bibinfo {year} {2009})}\BibitemShut {NoStop}%
\bibitem [{\citenamefont {Yan}\ \emph {et~al.}(2022)\citenamefont {Yan},
  \citenamefont {Revesz}, \citenamefont {Liang},\ and\ \citenamefont
  {Li}}]{Yan2022}%
  \BibitemOpen
  \bibfield  {author} {\bibinfo {author} {\bibfnamefont {J.}~\bibnamefont
  {Yan}}, \bibinfo {author} {\bibfnamefont {S.}~\bibnamefont {Revesz}},
  \bibinfo {author} {\bibfnamefont {D.}~\bibnamefont {Liang}}, \ and\ \bibinfo
  {author} {\bibfnamefont {H.}~\bibnamefont {Li}},\ }\href {\doibase
  10.1103/PhysRevA.105.052810} {\bibfield  {journal} {\bibinfo  {journal}
  {Phys. Rev. A}\ }\textbf {\bibinfo {volume} {105}},\ \bibinfo {pages}
  {052810} (\bibinfo {year} {2022})}\BibitemShut {NoStop}%
\bibitem [{\citenamefont {Myers}\ \emph {et~al.}(2010)\citenamefont {Myers},
  \citenamefont {Lewis}, \citenamefont {Fuller}, \citenamefont {Tekavec},
  \citenamefont {Yocum},\ and\ \citenamefont {Ogilvie}}]{Myers2010}%
  \BibitemOpen
  \bibfield  {author} {\bibinfo {author} {\bibfnamefont {J.~A.}\ \bibnamefont
  {Myers}}, \bibinfo {author} {\bibfnamefont {K.~L.~M.}\ \bibnamefont {Lewis}},
  \bibinfo {author} {\bibfnamefont {F.~D.}\ \bibnamefont {Fuller}}, \bibinfo
  {author} {\bibfnamefont {P.~F.}\ \bibnamefont {Tekavec}}, \bibinfo {author}
  {\bibfnamefont {C.~F.}\ \bibnamefont {Yocum}}, \ and\ \bibinfo {author}
  {\bibfnamefont {J.~P.}\ \bibnamefont {Ogilvie}},\ }\href {\doibase
  10.1021/jz100972z} {\bibfield  {journal} {\bibinfo  {journal} {The Journal of
  Physical Chemistry Letters}\ }\textbf {\bibinfo {volume} {1}},\ \bibinfo
  {pages} {2774} (\bibinfo {year} {2010})}\BibitemShut {NoStop}%
\bibitem [{\citenamefont {Thämer}\ \emph {et~al.}(2015)\citenamefont
  {Thämer}, \citenamefont {Marco}, \citenamefont {Ramasesha}, \citenamefont
  {Mandal},\ and\ \citenamefont {Tokmakoff}}]{Martin2015}%
  \BibitemOpen
  \bibfield  {author} {\bibinfo {author} {\bibfnamefont {M.}~\bibnamefont
  {Thämer}}, \bibinfo {author} {\bibfnamefont {L.~D.}\ \bibnamefont {Marco}},
  \bibinfo {author} {\bibfnamefont {K.}~\bibnamefont {Ramasesha}}, \bibinfo
  {author} {\bibfnamefont {A.}~\bibnamefont {Mandal}}, \ and\ \bibinfo {author}
  {\bibfnamefont {A.}~\bibnamefont {Tokmakoff}},\ }\href {\doibase
  10.1126/science.aab3908} {\bibfield  {journal} {\bibinfo  {journal}
  {Science}\ }\textbf {\bibinfo {volume} {350}},\ \bibinfo {pages} {78}
  (\bibinfo {year} {2015})}\BibitemShut {NoStop}%
\bibitem [{\citenamefont {Xiang}\ \emph {et~al.}(2020)\citenamefont {Xiang},
  \citenamefont {Ribeiro}, \citenamefont {Du}, \citenamefont {Chen},
  \citenamefont {Yang}, \citenamefont {Wang}, \citenamefont {Yuen-Zhou},\ and\
  \citenamefont {Xiong}}]{Xiang2020}%
  \BibitemOpen
  \bibfield  {author} {\bibinfo {author} {\bibfnamefont {B.}~\bibnamefont
  {Xiang}}, \bibinfo {author} {\bibfnamefont {R.~F.}\ \bibnamefont {Ribeiro}},
  \bibinfo {author} {\bibfnamefont {M.}~\bibnamefont {Du}}, \bibinfo {author}
  {\bibfnamefont {L.}~\bibnamefont {Chen}}, \bibinfo {author} {\bibfnamefont
  {Z.}~\bibnamefont {Yang}}, \bibinfo {author} {\bibfnamefont {J.}~\bibnamefont
  {Wang}}, \bibinfo {author} {\bibfnamefont {J.}~\bibnamefont {Yuen-Zhou}}, \
  and\ \bibinfo {author} {\bibfnamefont {W.}~\bibnamefont {Xiong}},\ }\href
  {\doibase 10.1126/science.aba3544} {\bibfield  {journal} {\bibinfo  {journal}
  {Science}\ }\textbf {\bibinfo {volume} {368}},\ \bibinfo {pages} {665}
  (\bibinfo {year} {2020})}\BibitemShut {NoStop}%
\bibitem [{\citenamefont {Song}\ \emph {et~al.}(2021)\citenamefont {Song},
  \citenamefont {Sechrist}, \citenamefont {Nguyen}, \citenamefont {Johnson},
  \citenamefont {Abramavicius}, \citenamefont {Redding},\ and\ \citenamefont
  {Ogilvie}}]{Song2021}%
  \BibitemOpen
  \bibfield  {author} {\bibinfo {author} {\bibfnamefont {Y.}~\bibnamefont
  {Song}}, \bibinfo {author} {\bibfnamefont {R.}~\bibnamefont {Sechrist}},
  \bibinfo {author} {\bibfnamefont {H.~H.}\ \bibnamefont {Nguyen}}, \bibinfo
  {author} {\bibfnamefont {W.}~\bibnamefont {Johnson}}, \bibinfo {author}
  {\bibfnamefont {D.}~\bibnamefont {Abramavicius}}, \bibinfo {author}
  {\bibfnamefont {K.~E.}\ \bibnamefont {Redding}}, \ and\ \bibinfo {author}
  {\bibfnamefont {J.~P.}\ \bibnamefont {Ogilvie}},\ }\href {\doibase
  10.1038/s41467-021-23060-9} {\bibfield  {journal} {\bibinfo  {journal}
  {Nature Communications}\ }\textbf {\bibinfo {volume} {12}},\ \bibinfo {pages}
  {2801} (\bibinfo {year} {2021})}\BibitemShut {NoStop}%
\bibitem [{\citenamefont {Hamm}(1995)}]{Hamm1995}%
  \BibitemOpen
  \bibfield  {author} {\bibinfo {author} {\bibfnamefont {P.}~\bibnamefont
  {Hamm}},\ }\href {\doibase https://doi.org/10.1016/0301-0104(95)00262-6}
  {\bibfield  {journal} {\bibinfo  {journal} {Chemical Physics}\ }\textbf
  {\bibinfo {volume} {200}},\ \bibinfo {pages} {415} (\bibinfo {year}
  {1995})}\BibitemShut {NoStop}%
\bibitem [{\citenamefont {Brosseau}\ \emph {et~al.}(2023)\citenamefont
  {Brosseau}, \citenamefont {Seiler}, \citenamefont {Palato}, \citenamefont
  {Sonnichsen}, \citenamefont {Baker}, \citenamefont {Socie}, \citenamefont
  {Strandell},\ and\ \citenamefont {Kambhampati}}]{Patrick2023}%
  \BibitemOpen
  \bibfield  {author} {\bibinfo {author} {\bibfnamefont {P.}~\bibnamefont
  {Brosseau}}, \bibinfo {author} {\bibfnamefont {H.}~\bibnamefont {Seiler}},
  \bibinfo {author} {\bibfnamefont {S.}~\bibnamefont {Palato}}, \bibinfo
  {author} {\bibfnamefont {C.}~\bibnamefont {Sonnichsen}}, \bibinfo {author}
  {\bibfnamefont {H.}~\bibnamefont {Baker}}, \bibinfo {author} {\bibfnamefont
  {E.}~\bibnamefont {Socie}}, \bibinfo {author} {\bibfnamefont
  {D.}~\bibnamefont {Strandell}}, \ and\ \bibinfo {author} {\bibfnamefont
  {P.}~\bibnamefont {Kambhampati}},\ }\href {\doibase 10.1063/5.0138252}
  {\bibfield  {journal} {\bibinfo  {journal} {The Journal of Chemical Physics}\
  }\textbf {\bibinfo {volume} {158}},\ \bibinfo {pages} {084201} (\bibinfo
  {year} {2023})}\BibitemShut {NoStop}%
\bibitem [{\citenamefont {Steck}()}]{Steck}%
  \BibitemOpen
  \bibfield  {author} {\bibinfo {author} {\bibfnamefont {D.~A.}\ \bibnamefont
  {Steck}},\ }\href@noop {} {\enquote {\bibinfo {title} {Rubidium 87 d line
  data},}\ }\Eprint {http://arxiv.org/abs/http://steck.us/alkalidata}
  {http://steck.us/alkalidata} \BibitemShut {NoStop}%
\end{thebibliography}%

\end{document}